\documentclass[12pt,a4paper]{article}
\pdfoutput=1

\usepackage{jheppub}
\usepackage{amsmath,amssymb,amsfonts,mathrsfs,mathtools,graphicx,bbm}
\usepackage{fixltx2e}
\usepackage[font=small]{caption}
\usepackage{lmodern}
\usepackage{booktabs}
\usepackage{dsfont}
\usepackage{slashed}
\usepackage{braket}
\usepackage{xspace}

\usepackage{epsfig}
\usepackage{relsize}
\input epsf
\usepackage{enumitem}
\usepackage{color}

\usepackage{hyperref}

\newcommand{\be}{\begin{equation}}
\newcommand{\ee}{\end{equation}}
\newcommand{\bea}{\begin{eqnarray}}
\newcommand{\eea}{\end{eqnarray}}
\newcommand{\bee}{\begin{enumerate}}
\newcommand{\eee}{\end{enumerate}}
\newcommand{\bei}{\begin{itemize}}
\newcommand{\eei}{\end{itemize}}
\newcommand{\bal}{\begin{equation}\begin{aligned}}
\newcommand{\eal}{\end{aligned}\end{equation}}
\newcommand{\bem}{\left (\begin{matrix}}
\newcommand{\eem}{\end{matrix} \right )}

\newcommand{\la}{\label}

\newcommand{\alg}[1]{\mathfrak{#1}}

\newcommand{\un}{\alg{u}}
\def\sl2{\alg{sl}(2)}

\newcommand{\ads}{${\rm  AdS}_5\times {\rm S}^5\ $}

\numberwithin{equation}{section}
\makeatletter
 \let\old@startsection=\@startsection
 \let\oldl@section=\l@section
 \renewcommand{\@startsection}[6]{\old@startsection{#1}{#2}{#3}{#4}{#5}{#6\mathversion{bold}}}
 \renewcommand{\l@section}[2]{\oldl@section{\mathversion{bold}#1}{#2}}
\makeatother
\DeclareMathOperator{\diag}{diag}

\DeclareMathOperator{\const}{const}

\newcommand{\AdS}{\text{AdS}}

\definecolor{grey}{rgb}{0.4,0.4,0.5}
\definecolor{darkgreen}{rgb}{0,0.5,0}
\definecolor{darkred}{rgb}{0.6,0.0,0}
\definecolor{lightbrown}{rgb}{1,0.9,0.8}
\definecolor{brown}{rgb}{0.6,0.3,0.3}
\definecolor{darkblue}{rgb}{0,0,0.5}
\definecolor{darkmagenta}{rgb}{0.5,0,0.5}

\definecolor{TCDblue}{rgb}{0.1,0.3,0.6}

\def\pa {\partial}
\def\ov{\over}

\def\a {\alpha}
\def\b {\beta}
\def\g {\gamma}

\def\de {\delta}

\def\e{\epsilon}

\def\k{\kappa}

\def\m{\mu}
\def\r {\rho}

\def\s {\sigma}

\def\z {\zeta}
\def\w {\omega}

\def\p{\phi}

\def\vp{\varphi}

\def\u\upsilon
\def\U\Upsilon
\def\vU\varUpsilon
\def\vPi{\varPi}

\def\cH{{\cal H}}

\def\cJ{{\cal J}}

\def\cL{{\cal L}}

\def\cO{{\cal O}}

\def\cT{{\cal T}}

\def\bR{{\mathbb R}}

\def\rK{{\rm K}}

\def\rV{{\rm V}}

\def\dX{{\dot X}}

\def\th{{\tilde h}}

\def\tw{{\tilde w}}

\def\x'{\mathaccent 19 x}
\def\y'{\mathaccent 19 y}
\def\n'{\mathaccent 19 n}
\def\u'{\mathaccent 19 u}

\def\et'{\mathaccent 19 \eta}
\def\th'{\mathaccent 19 \theta}
\def\lam'{\mathaccent 19 \lambda}
\def\varet'{\mathaccent 19 \vartheta}
\def\rh'{\mathaccent 19 \rho}
\def\xb'{\mathaccent 19 {\bar{x}}}

\def\Gt{{\widetilde G}}
\def\Bt{{\widetilde B}}

\def\d1{{\dot{1}}}

\def\sun2{SU(N)$\times$SU(N)}

\def\TTb{${T\overline{T}}$}

\def\ph{{\rm{ph}}}

\newcommand{\psib}{\bar{\psi}}
\newcommand{\psid}{\dot{\psi}}
\newcommand{\psibd}{\dot{\psib}}
\newcommand{\Ab}{\bar{A}}

\newcommand{\psip}{{\psi'}}
\newcommand{\psibp}{{\psib'}}

\newcommand{\rhod}{\dot{\rho}}
\newcommand{\rhop}{{\rho'}}
\newcommand{\phid}{\dot{\phi}}
\newcommand{\phip}{{\phi'}}

\newcommand{\bpm}{\begin{pmatrix}}
\newcommand{\epm}{\end{pmatrix}}

\title{\boldmath $T\overline{T}$ deformations of non-relativistic models}

 \author{Chantelle Esper }
\author{and  Sergey Frolov\footnote{ Correspondent fellow at
Steklov Mathematical Institute, Moscow.}}
\affiliation{School of Mathematics and Hamilton Mathematics Institute, \\
Trinity College, Dublin 2,
Ireland}

\emailAdd{esperc@tcd.ie}
\emailAdd{frolovs@maths.tcd.ie}
\abstract{The light-cone gauge approach to $T\overline{T}$ deformed models is used to derive the $T\overline{T}$  deformed matrix nonlinear Schr\"odinger equation, the Landau--Lifshitz equation, and the Gardner equation.
Properties of one-soliton solutions of the $T\overline{T}$ deformed nonlinear Schr\"odinger and Korteweg--de Vries equations are discussed in detail. The NLS soliton  exhibits the recently discussed phenomenon of widening/narrowing width of particles under the $T\overline{T}$ deformation. However, whether the soliton's size is increasing or decreasing depends not only on the sign of the deformation parameter but also on soliton and potential parameters. The $T\overline{T}$ deformed KdV equation admits a one-parameter family of one-soliton solutions in addition to the usual velocity parameter. 
The extra parameter modifies the properties of the soliton, in particular, it appears in the dispersion relation. }

\begin{document}

\null\vskip-40pt
 \vskip-5pt \hfill
\vskip-5pt \hfill {\tt\footnotesize
TCDMATH 21-03}
% \vskip-5pt \hfill {\tt\footnotesize HMI-11-05}

\maketitle
\flushbottom

%\newpage

%\tableofcontents

\renewcommand{\thefootnote}{\arabic{footnote}}
\setcounter{footnote}{0}

%%%%%%%%%%%%%%%%%%%%%%%%%%%%%%%%%%%%%%%%%%
\section{Introduction }
%%%%%%%%%%%%%%%%%%%%%%%%%%%%%%%%%%%%%%%%%%

The irrelevant \TTb\  deformation of  two-dimensional Lorentz invariant models
introduced in \cite{Z04} has many interesting properties. In particular, if a seed model is integrable then the \TTb\ deformed model is also integrable at least at the classical level  \cite{SZ16, Tateo16}. Assuming the \TTb\ operator is well-defined at the quantum level, the factorisation of two-point correlation functions at large separation  and a CFT limit at short distances,  one can show that the spectrum of a \TTb\ deformed model is governed by an inhomogeneous inviscid Burgers equation.  If the spectrum depends regularly on the deformation parameter then it is  completely fixed by the spectrum of the seed model \cite{Z04}. The Burgers equation can be used to derive the CDD factor which relates the S-matrices of the deformed and seed models \cite{SZ16}. The same CDD factor  appears in the world-sheet S-matrix of the  light-cone gauge-fixed \ads string sigma model \cite{AFZ06b} and in the study of effective bosonic string theory in flat space \cite{Dubovsky12}. It also describes the world-sheet scattering of light-cone strings on AdS$_3$ backgrounds without RR fields
\cite{Sfondrini18, Sfondrini18a,Sfondrini18c}.  Its relation to the \TTb\ deformation was pointed out in \cite{Tateo13}.  For many other aspects of \TTb\ deformed models see the lecture notes  \cite{Jiang19}. 

There are various connections of \TTb\ deformed relativistic models to two-dimensional gravity. 
A \TTb\ deformed S-matrix and the partition function can be obtained by coupling a seed model to the flat space Jackiw-Teitelboim (JT)  gravity and its generalisations \cite{Dubovsky17,Dubovsky18,Tolley}.  This leads to the interpretation of  the \TTb\ deformation as a nonlocal field dependent change of space-time coordinates of the seed model \cite{Tateo18b}. The partition function of a deformed model can also be derived by coupling a seed model to a random geometry \cite{Cardy2018a}. 
The action of a \TTb\ deformed model can be obtained by interpreting it \cite{Sfondrini18b,SF19a}  as the action 
of a non-critical string sigma model in a parameter dependent uniform light-cone gauge introduced in \cite{AFZ06a}. 
Most of the \TTb\ deformed Lorentz invariant actions \cite{AAF,Tateo16,Bonelli18,Tateo18a,Sfondrini18b,Sfondrini19a,Sethi18,Sfondrini19b,Freedman19a,Ouyang20,Chakrabarti20,Aghbolah20} derived by using other methods are particular cases of the \TTb\ deformed action for a very general system of any number of bosons and fermions  with an arbitrary potential which was derived in \cite{SF19a} by using the light-cone gauge approach. In fact, for \TTb\ deformations with the canonical stress-energy tensor this action is universal and can be applied to any model.

The \TTb\ deformation of non-Lorentz invariant models is also very interesting to study even at the classical level. 
Many non-relativistic models, for example the nonlinear Schr\"odinger (NLS) equation, the Landau-Lifshitz (LL) equation and the Gardner equation which is a combination of the Korteweg--de\,Vries (KdV) and the modified KdV (mKdV) equation,  play important roles in describing various phenomena in nonlinear optics, hydrodynamics, plasma physics and condensed matter physics. Some aspects of  non-Lorentz invariant \TTb\ deformed models have been studied in 
\cite{Cardy2018b,Cardy2020, Jiang2020a,Jiang2020b,Tateo2020,Chen2020}. 

The light-cone gauge approach to \TTb\ deformed models works equally well for relativistic and non-relativistic models.
 In particular, as was mentioned in \cite{SF19a}, it could be used to derive the \TTb\ deformed action for the chiral SYK model and the matrix nonlinear Schr\"odinger model. 
 
In this paper we derive the \TTb\ deformed actions for the matrix NLS equation, the LL equation and the Gardner equation by using the light-cone gauge approach. The resulting actions are written in the first-order form and depend on auxiliary fields. For the deformed matrix NLS and  LL models, the auxiliary fields satisfy algebraic equations of motion and can be eliminated  leading to Nambu-Goto type actions.  The \TTb\ deformed Gardner model is more involved because the auxiliary fields appear in  the deformed action together with their space derivatives, and it is unlikely that there exists a local deformed action depending only on the physical field. Moreover, the Gardner field which appears in the Gardner equation is not the physical field of the Gardner model action  but one of the auxiliary fields. 

We then find one-soliton solutions of the deformed NLS and KdV  models. The deformed NLS soliton clearly exhibits the general phenomenon of widening/narrowing the width of particles under the \TTb\ deformation recently discussed in \cite{Cardy2020}. However, in the nonrelativistic case whether the soliton's size is increasing or decreasing depends not only on the sign of the deformation parameter but also on soliton and potential parameters. As to the \TTb\ deformed KdV soliton, we find a one-parameter family of solutions where the extra parameter is related to  the time dependence of the physical field at space infinities. If one fixes the dependence, then the extra parameter can be interpreted as the parameter of the deformation by the time component of the conserved current due to the invariance of the \TTb\ deformed Gardner model under constant shifts of its physical field. The parameter modifies the properties of the soliton, in particular, it appears in the dispersion relation.
 All these solutions reduce to the usual KdV soliton once one takes the \TTb\ deformation parameter to 0.

The \TTb\ deformed action for the (non-matrix) NLS model has been also found in \cite{Jiang2020b,Tateo2020,Chen2020} by using  different and substantially more complicated methods than the light-cone gauge one, and some deformed soliton solutions  have been analysed in \cite{Tateo2020}. 

The paper is organised as follows. In section 2 we first review the universal \TTb\ deformed action derived in \cite{SF19a} and introduce our notations. Then in section 2.1, as a warm-up, we obtain the well-known \TTb\ deformed Lagrangian of a sigma-model of scalar fields with arbitrary potential and $B$-field. In section 2.2-2.4 we get the \TTb\ deformed actions for the  matrix NLS, the LL and the Gardner models. In section 2.3 we also show how the deformed NLS and sine-Gordon models can be obtained from the deformed LL model by taking appropriate limits generalising the well-known results for the seed models \cite{faddeev}. In section 3.1 we discuss a one-soliton solution of the \TTb\ deformed NLS equation with the potential which in addition to the usual quartic term also includes the density of particles. This term is unimportant for the undeformed NLS model because it can be removed by a time dependent U(1) transformation of the fields. The \TTb\ deformed model and its solutions however depend  on it in a  nontrivial way. In section 3.2 we consider a one-parameter family of one-soliton solutions of the \TTb\ deformed KdV equation which is the simplest case of the Gardner equation.    In Conclusions we summarise the results obtained and discuss numerous open problems. Technical details can be found in several Appendices.

%\newpage

%%%%%%%%%%%%%%%%%%%%%%%%%%%%%%%%%%%%%%%%%%
\section{Lagrangians of  \texorpdfstring\TTb\ \  deformed models}\la{lcg}
%%%%%%%%%%%%%%%%%%%%%%%%%%%%%%%%%%%%%%%%%%

All models we are going to discuss in this paper are \TTb\ deformations of a seed model described by the
following action
\bal\la{Sseed} 
S_0 = \int\, {\rm
d}x{\rm d}t\, \cL_0\,,\quad \cL_0=P_a^t(\Psi)\pa_t\Psi^a+P_a^x(\Psi)\pa_x\Psi^a-\rV(\Psi)\,.
\eal
Here $\Psi^a\,,\, a=1,\ldots, n$ are bosonic and fermionic fields which can be real or complex. If a field is complex then the set $(\Psi^a)$ also includes its complex conjugate field.  $P_{a}^t$, $P_{a}^x$ and $V$ are chosen so that the action \eqref{Sseed}  is real and Grassmann even but otherwise they are arbitrary functions of the fields  $\Psi^a$. The seed action is written in the first-order formalism with respect to both time and space, and as a result many of the fields are non-dynamical. If each $\Psi^a$ belongs to a Lorentz group representation and $P_{a}^t$, $P_{a}^x$ belong to the conjugate representation, and $V$ is a Lorentz scalar  then the seed model is Lorentz invariant.

The light-cone gauge approach to \TTb\ deformed models developed in \cite{SF19a} then leads to the following deformed Lagrangian
\bal\la{Lttb}
\cL={\rK^t_t +\rK^x_x- \rV +\a(\rK^t_t\rK^x_x-\rK^t_x\rK^x_t) \ov 1+\a\rV } = {\cL_0 -{\a\ov2}\e^{\g\r}\e_{\mu\nu}\rK^\mu_\g\rK^\nu_\r\ov 1+\a\rV }\,,
\eal 
where
\bal
\rK^t_\g\equiv P_a^t\pa_\g\Psi^a\,,\quad \rK^x_\g\equiv P_a^x\pa_\g\Psi^a\,,\quad \g=t,x\,,
\eal
and the skew-symmetric Levi-Civita symbol is defined by  $\e^{01}=\e^{tx}=1=\e_{xt}=\e_{10}$.
To get \eqref{Lttb} from the Lagrangian (3.53) in \cite{SF19a} one should make the following replacements in (3.53): $\Psi_\pm\to\Psi$, $\Psi^aK^+_{ab}\to -i\, P^t_{b}$, $\Psi^aK^-_{ab}\to -i\,P^x_{b}$, $\pa_+\to\pa_t$, $\pa_-\to\pa_x$.%, $\s\to x, \tau\to t$. 

The canonical stress-energy tensor of the deformed model can be easily calculated
\bal
T^\mu{}_\nu=  {\pa\cL\ov \pa\pa_\mu\Psi^a}\pa_\nu\Psi^a -\de^\mu_\nu\cL
\eal
\bal
T^t{}_t={-\rK^x_x+\rV\ov 1+\a \rV} \,,\quad T^x{}_t={\rK^x_t\ov 1+\a \rV} \,,
\quad
T^t{}_x={\rK^t_x\ov 1+\a \rV} \,,\quad T^x{}_x={-\rK^t_t+\rV\ov 1+\a \rV} \,,
\eal
and used to check that the deformed Lagrangian \eqref{Lttb} satisfies the flow equation
\bal
{\pa\cL\ov\pa\a} = T^t{}_t T^x{}_x - T^t{}_xT^x{}_t
\eal
Since any seed model can be written in the form \eqref{Sseed}, the \TTb\ deformed Lagrangian \eqref{Lttb}  is universal. However, in a non-relativistic case the seed Lagrangian \eqref{Sseed} may also include total derivative terms which do not change the equations of motion of the seed model but they do change the canonical stress-energy tensor and as a result the Lagrangian and the equations of motion of the deformed model may depend on the total derivative terms. 
This dependence does not seem to be spurious, and we do not think that it can be undone by a field redefinition. 

%%%%%%%%%%%%%%%%%%%%%%%%%%%%%%%%%%%%%%%%%%
\subsection{\texorpdfstring\TTb\  \  deformed sigma model}\la{sigma}
%%%%%%%%%%%%%%%%%%%%%%%%%%%%%%%%%%%%%%%%%%

As a warm-up, in this subsection we discuss the well-known deformation of a sigma-model  of $n$ scalar fields described by the Lagrangian
\bal\la{Lbos} 
\cL_0 ={1\over 2}\eta^{\a\b}\partial_\a X^i\partial_\b X^j\,G_{ij}(X)+{1\over 2}\e^{\a\b}\partial_\a X^i\partial_\b X^j\,B_{ij}(X)-U(X)\,, 
\eal
where $\eta^{\a\b}=\diag(1,-1)$, $\e^{01}=\e^{tx}=1=\e_{xt}$,  and $U$ is an arbitrary potential. 

To bring the Lagrangian to the form \eqref{Sseed}, we
 introduce the momentum vectors
\bal
P^\a_i={\pa \cL_0\ov\pa\pa_\a X^i} =\big(\eta^{\a\b}\,G_{ij}+\e^{\a\b}\,B_{ij} \big)\partial_\b X^j\,.
\eal
The component $P^t_i$ is the momentum conjugate to $X^i$.

Solving these equations for $\pa_\a X^i$, one finds
\bal
\pa_\a X^i = \big(\eta_{\a\b}\, \Gt^{ij}+\e_{\a\b}\,\Bt^{ij} \big)P^\b_j\,,
\eal
where $ \Gt^{ij}$ and $ \Bt^{ij}$ satisfy
\bal
G_{ij}\Gt^{jk}+B_{ij}\Bt^{jk} =\de_i^k\,,\quad G_{ij}\Bt^{jk}+B_{ij}\Gt^{jk} =0\,,
\eal
which can be solved as
\bal
 \Gt^{ij}\big(G_{jk}-B_{jl} G^{lm}B_{mk} \big)=\de^i_k\,,\quad  \Bt^{ij}=- \Gt^{ik}B_{kl}G^{lj}=- G^{ik}B_{kl}\Gt^{lj}\,.
\eal
Note that $\Gt$ is symmetric and $\Bt$ is anti-symmetric.

It is then straightforward to
 rewrite $\cL_0$ in the first-order formalism
\bal\la{Lbos2} 
\cL_0 = P^\g_i\pa_\g X^i- {1\over 2}\big(\eta_{\g\r}\,\Gt^{ij}+\e_{\g\r}\,\Bt_{ij}\big)P^\g_iP^\r_j  -U\,.
\eal
It is the form of $\cL_0$ we need. The set $(\Psi^a)$ consists of $X^i$, and $P_i^\g$, and 
\bal
\rK^t_t &= P^t_i\pa_t X^i\,,\quad \rK^x_x = P^x_i\pa_x X^i\,,\quad \rK^t_x = P^t_i\pa_x X^i\,,\quad \rK^x_t = P^x_i\pa_t X^i\,, 
\\
V &={1\over 2}\big(\eta_{\g\r}\,\Gt^{ij}+\e_{\g\r}\,\Bt_{ij}\big)P^\g_iP^\r_j +U\,.
\eal
Thus, the \TTb\ deformed Lagrangian of the sigma model is
\bal\la{Lttbsigma}
\cL={P^\g_i\pa_\g X^i- {1\over 2}\big(\eta_{\g\r}\,\Gt^{ij}+\e_{\g\r}\,\Bt_{ij}\big)P^\g_iP^\r_j  -U -{\a\ov2}\e^{\g\r}\e_{\mu\nu}P^\mu_i\pa_\g X^iP^\nu_j\pa_\r X^j \ov 1+{\a\over 2}\big(\eta_{\g\r}\,\Gt^{ij}+\e_{\g\r}\,\Bt_{ij}\big)P^\g_iP^\r_j +\a U }\,.
\eal 
One can get rid of the auxiliary fields $P_i^\g$ by using their equations of motion and, choosing a proper solution of the resulting quadratic equation on $\cL$, one gets the well-known answer\footnote{To find $\cL_\ph$ which depends only on the physical fields $X^i$ it is not necessary to solve the equations of motion for $P_i^\a$. Since $\cL$ depends just on $K^\g_\r$ and $V$ it is sufficient to know only them to find $\cL_\ph$. This can be done by expressing $V$ in terms of $\cL$ and $K^\g_\r$, and substituting it into the equations of motion for $P_i^\a$. This leads to simple linear equations for $K^\g_\r$ which can be easily solved. The consistency condition of the solution with the expression for $V$ in terms of $\cL$ and $K^\g_\r$ leads to a quadratic equation for $\cL$  with coefficients which depend only on the physical fields.}
 \bal\la{LbosTTb}
\cL_\ph=-{1\ov\a}+{1\ov 2\tilde\a}+{1\ov2\tilde\a}\sqrt{1+2\tilde\a(\dX^2-X'^2) -4\tilde\a^2(\dX^2 X'^2-(\dX X')^2)} +\dX^iX'^j\,B_{ij}\,,
\eal
where
\bal
\dX^2\equiv  G_{ij}\dX^i\dX^j\,,\quad X'^2\equiv  G_{ij}X'^iX'^j\,,\quad \dX X'\equiv  G_{ij}\dX^iX'^j\,,\quad \tilde\a = \a(1 +\a U)\,.
\eal
It is worth stressing that the Lagrangian \eqref{Lttbsigma} describes both the perturbative and non-perturbative in $\a$ solutions of the quadratic equation on $\cL$.

%%%%%%%%%%%%%%%%%%%%%%%%%%%%%%%%%%%%%%%%%%
\subsection{\texorpdfstring\TTb\  \ deformed matrix nonlinear  Schr\"odinger model}\la{nlse}
%%%%%%%%%%%%%%%%%%%%%%%%%%%%%%%%%%%%%%%%%%

The Lagrangian of the matrix nonlinear  Schr\"odinger model is
\bal\la{LmNS}
\cL_0 = \frac{i}{2}(\psib \psid - \psibd \psi) - \psib' \psip -U\,,\quad U=\kappa\, \psib \psi\psib \psi - \mu\,\psib \psi\,.
\eal
Here 
\bal
\psi = (\psi_{ai})\,,\quad  \psib =\psi^\dagger= (\psi_{ia}^*)\,, \quad a=1,\ldots,n\,,\, i=1,\ldots,m
\eal 
are  complex $n\times m$ and $m\times n$ matrices hermitian conjugate to each other. Then, the trace is implied in \eqref{LmNS}, i.e.
\bal
\psib \psid \equiv \psi_{ia}^* \psid_{ai}\,,\quad  \psib \psi\psib \psi\equiv \psi_{ia}^* \psi_{aj}\psi_{jb}^* \psi_{bi}\,.
\eal
To bring the Lagrangian into the desired form we introduce two auxiliary matrices hermitian conjugate to each other
\bal
A = (A_{ai})\,,\quad  \Ab =A^\dagger= (A_{ia}^*)\,, \quad a=1,\ldots,n\,,\, i=1,\ldots,m\,,
\eal 
and rewrite \eqref{LmNS} as 
\bal\la{LmNS2}
\cL_0= \frac{i}{2}(\psib \psid - \psibd \psi) - \bar{A}\psi' -  \psib' A+ \bar{A}A - U
\eal
Thus, the set $(\Psi^a)$ consists of $\psi, \psib, A, \Ab$, and 
\bal
\rK^t_t &=  \frac{i}{2}(\psib \psid - \psibd \psi) \,,\quad \rK^x_x = - \bar{A}\psi' -  \psib' A\,,\\
 \rK^t_x &=  \frac{i}{2}(\psib \psi' - \psib' \psi) \,,\quad \rK^x_t = - \bar{A}\psid -  \psibd A\,, 
\quad
V =U-\bar{A}A\,,
\eal
where the trace is implied.

The \TTb\ deformed Lagrangian of the matrix nonlinear Schr\"odinger  model, therefore, is
\bal\la{Lttbnmsc}
\cL= \frac{K^t_t - \bar{A}\psi' -  \psib' A+ \bar{A}A - U - \alpha\big(K^t_t(\bar{A}\psi' +  \psib' A) - K^t_x(\bar{A}\psid +  \psibd A) \big)}{1-\alpha(\bar{A}A - U)}\,.
\eal 
Eliminating  the auxiliary fields $A, \Ab$ by using their equations of motion and, choosing the regular in $\a$ solution of the resulting quadratic equation on $\cL$, one gets 
 \bal\la{LTTbmnsc}
\cL_\ph&=-{1\ov\a}+{1+\a K^t_t+\sqrt{\Lambda} \ov 2\tilde\a}\,,\quad \tilde\a = \a(1 +\a U)\,,
   \\
   \Lambda&=(1+\a K^t_t)^2 (1-4 \tilde\a \psibp\psip)+4 \a \tilde\a (1+\a K^t_t)K^t_x 
   (\psibd\psip+\psibp\psid)-4 \a^2\, \tilde\a\, (K^t_x)^2 \,\psibd\psid
\eal
where in the expression for $\Lambda$ the trace is implied.

It is clear that the deformation drastically modifies the Poisson structure of the model, and developing a Hamiltonian formulation requires dealing with an intricate system of second-class constraints. The same seems to be valid for any non-relativistic model.

%%%%%%%%%%%%%%%%%%%%%%%%%%%%%%%%%%%%%%%%%%
\subsection{\texorpdfstring\TTb\ \ deformed Landau--Lifshitz model}\la{lle}
%%%%%%%%%%%%%%%%%%%%%%%%%%%%%%%%%%%%%%%%%%

We mostly follow the notations in \cite{faddeev}.

The Landau-Lifshitz equation is
\bal\la{LLeq}
{\pa S_i\ov \pa t} = {1\ov R^2}\e_{ijk}\,S_j{\pa^2 S_k\ov \pa x^2} + \e_{ijk}\,S_j J_kS_k \,,
\eal
where $S_i^2=R^2$, and we sum over repeated indices even if there are 3 of them. 

The fields  $S_i$ have the Poisson structure
\bal
\{ S_i(x),S_j(y)\}=-\eta\,\e_{ijk}\, S_k(x)\de(x-y)\,,
\eal
and the LL equation follows from the Hamiltonian
\bal
H = \int dx\, {1\ov2\eta}\Big({1\ov R^2}\Big( {\pa S_k\ov \pa x}\Big)^2 -  J_kS_k^2 +J_3R^2 \Big) \,.
\eal
where the constant $J_3R^2$ guaranties the vanishing of the Hamiltonian density 
 in the rapidly decreasing case where we impose the conditions $S_k(\pm\infty)=\de_{k3}R$.
By rescaling $S_k$ and $x,y$ one can set $R=1$ and $\eta=1$. We prefer to  keep these two parameters to simplify taking the limits to the NLS and sine-Gordon models.

To find the \TTb\ deformed LL model we need its Lagrangian description. To this end we multiply  \eqref{LLeq} by  $\e_{lmi}S_m$, and, changing the indices, get
\bal\la{LLeq2}
\e_{ijk}S_j{\pa S_k\ov \pa t} 
=- \big( \de_{ij}- {1\ov R^2}S_i S_j \big)\Big({\pa^2 S_j\ov \pa x^2}   +  R^2J_jS_j\Big)\,,\quad S_i^2=R^2\,.
\eal
These equations can be  derived from the following Lagrangian
\bal
L_0=\int dx\,\int_0^\infty dr\,{1\ov \eta R^2}{\e_{ijk}} S_i  {\pa S_j\ov \pa r} {\pa S_k\ov \pa t}+\int\, dx\,{1\ov2\eta}\Big(-{1\ov R^2}\Big( {\pa S_k\ov \pa x}\Big)^2 +  J_kS_k^2  -J_3R^2\Big)\,,
\eal
where $S_k$ are subject to the sphere constraint $S_k^2=R^2$. In the first term $S_k$ depend on an extra radial coordinate $r$, and satisfy the conditions $S_k(x,t,r)|_{r=0} = S_k(x,t)$, $S_k(x,t,r)|_{r=\infty} = \de_{k3} R$.  This is a WZNW type term, and its variation is
\bal
\de \int dx\,\int_0^\infty dr\,{1\ov \eta R^2}{\e_{ijk}} S_i  {\pa S_j\ov \pa r} {\pa S_k\ov \pa t}=-\int dx\,{\e_{ijk}\ov\eta R^2}  S_i \de S_j{\pa S_k\ov \pa t}\,,
\eal
where the variation $\de S_k$ is tangent to the sphere, i.e. it obeys the constraint  $\de S_k S_k=0$. Because of this, any products $ V_k \de S_k$ have to be replaced with $V_k(\de_{km}-{S_kS_m\ov R^2})\de S_k$. It produces all the  terms on the r.h.s. of  the equations of motion \eqref{LLeq2}. Introducing any coordinates $\p^a$, $a=1,2$ on the sphere $S_k^2=R^2$, one can bring the WZNW term to the total derivative form
  \bal
 {1\ov \eta R^2}{\e_{ijk}} S_i  {\pa S_j\ov \pa r} {\pa S_k\ov \pa t} =- {\pa\ov \pa r} \big(  P_a{\pa \p_a\ov \pa t} \big)+{\pa\ov \pa t} \big(  P_a{\pa \p_a\ov \pa r} \big)\,,
  \eal
where $P_a$ satisfies the condition $P_a(x,t,\infty)=0$ to ensure the absence of the contribution from the first term at $r=\infty$.
We will always drop the total time derivative term, integrate the remaining term over $r$ and, as a result, use the following Lagrangian (density) for the \TTb\ deformation
\bal
\cL_0=P_k \dot S_k-{1\ov 2\eta R^2}\Big( {\pa S_k\ov \pa x}\Big)^2 +  {1\ov2\eta}(J_kS_k^2-J_3R^2) -U_{\rm add}(S_k)\,,
\eal
where $P_k$ are such that $P_k \dot S_k=P_a\dot\p^a$, and $U_{\rm add}$ is an additional potential term which can be an arbitrary function of $S_k$.  We  will choose it later so that  the \TTb\ deformed NLS model could be obtained as a special limit of the \TTb\ deformed LL model.

In particular, in spherical coordinates 
\bal\la{sphcoord}
S_1=\cos\p\sin\theta\,,\quad S_2=\sin\p\sin\theta\,,\quad S_3=\cos\theta\,,
\eal
the WZNW term takes the form 
\bal
P_k \dot S_k= {1\ov \eta R^2}(\cos\theta-1)\dot\p = - {2\ov \eta R^2}\sin^2{\theta\ov2}\, \dot\p\,.
\eal

Now, introducing an auxiliary vector $A_i$, the LL model  Lagrangian  can be written as
\bal
\cL_0=P_k \dot S_k +A_k S_k'+{\eta R^2\ov 2}A_k^2  +{1\ov2\eta} (J_kS_k^2 -J_3R^2)-U_{\rm add}
\eal

We see that the set $(\Psi^a)$ consists of $S_k, A_k$, and 
\bal
\rK^t_t &= P_k\dot S_k\,,\quad
 \rK^t_x = P_k S'_k\,,\quad \rK^x_x = A_k S_k'\,,\quad \rK^x_t =A_k\dot S_k\,, 
\\
V &=-{\eta R^2\ov 2}A_k^2 +U\,,\quad U=  -{1\ov2\eta} (J_kS_k^2 -J_3R^2)+U_{\rm add}
\,.
\eal
Thus, the \TTb\ deformed Lagrangian of the LL model is
\bal\la{LttbLL}
\cL={P_k \dot S_k +A_k S_k'+{\eta R^2\ov 2}A_k^2  +{1\ov2\eta} (J_kS_k^2 -J_3R^2)-U_{\rm add}
+\a P_k A_l  (\dot S_k S_l'-S'_k\dot S_l) \ov 1-{\a\eta R^2\ov 2}A_k^2  -{\a\ov2\eta}(J_kS_k^2-J_3R^2) +\a U_{\rm add}}\,.
\eal 
One can get rid of the auxiliary fields $A_k$ by using their equations of motion and, choosing a proper solution of the resulting quadratic equation on $\cL$, one gets 
 \bal\la{LTTbLL}
\cL_\ph&=-{1\ov\a}+{1+\a K^t_t+\sqrt{\Lambda} \ov 2\tilde\a}\,,\quad \tilde\a = \a(1 +\a U)\,,
   \\
   \Lambda&=(1+\a K^t_t)^2 (1-{2\tilde\a\ov\eta R^2} S_k'^2)+{4 \a\ov\eta R^2}  \tilde\a (1+\a K^t_t)K^t_x 
 S_k'\dot S_k-{2 \a^2\tilde\a\ov\eta R^2} \, (K^t_x)^2 \,\dot S_k^2\,.
\eal

The similarity of this Lagrangian with \eqref{LTTbmnsc} for the NLS model is obvious, and not accidental. It is well-known that the NLS model can be obtained from the LL model  \cite{faddeev}. Since the NLS model has a U(1) symmetry we need to set $J_1=J_2=J$. Then, the LL model also has the symmetry and $S_3$ is proportional to  the density of the U(1) current, and it can be added to the LL Lagrangian while preserving the integrability of the model. Thus, the potential $U$ we are going to use is
\bal
U={1\ov2\eta}(J_3-J) (R^2-S_3^2) + \nu (R-S_3)\,,
\eal
where $\nu$ is any constant.

Next, we use the spherical coordinates \eqref{sphcoord}, and  get
\bal
\rK^t_t &= P_k\dot S_k= - {2\ov \eta R^2}\sin^2{\theta\ov2}\,\dot\p\,,\quad
 \rK^t_x = P_k S'_k= - {2\ov \eta R^2}\sin^2{\theta\ov2}\,\p'\,,\\
  {1\ov\eta R^2} S_k'^2&={1\ov\eta R^2} (\theta'^2+\sin^2\theta \p'^2)\,,\quad  {1\ov\eta R^2} \dot S_k^2={1\ov\eta R^2} (\dot\theta^2+\sin^2\theta \dot\p^2)\,,\\
  {1\ov\eta R^2} S_k'\dot S_k&={1\ov\eta R^2} (\theta\dot\theta+\sin^2\theta \p'\dot\p)\,,\quad 
  U={1\ov2\eta } ( J_3-J )\sin^2\theta +\nu (1-\cos\theta)\,.
\eal
Now,
we set $R=1$, and rescale the angle $\theta$ as
\bal
\theta= \sqrt{2\eta}\, \r\,.
\eal
We want to  take the limit $\eta\to 0$ and get a finite Lagrangian. We first obtain
\bal
\rK^t_t &\to -\r^2\dot\p\,,\quad
 \rK^t_x \to -\r^2\p'\,,\quad
  {1\ov\eta R^2} S_k'^2\to 2(\r'^2+\r^2 \p'^2)\,,\quad  
  \\
  {1\ov\eta R^2} \dot S_k^2&\to 2(\dot\r^2+\r^2 \dot\p^2)\,,\quad
  {1\ov\eta R^2} S_k'\dot S_k\to 2(\r'\dot\r+\r^2 \p'\dot\p)\,. 
 \eal
 To make contact with the NLS model, we introduce $\psi,\psib$ as
 \bal
 \psi=\r\, e^{i\phi}\,,\quad  \psib=\r\, e^{-i\phi}\,,
 \eal
 and find
 \bal
 -\r^2\dot\p&= \frac{i}{2}(\psib \psid - \psibd \psi)  \,,\quad -\r^2\p'=\frac{i}{2}(\psib \psi' - \psib' \psi) \,,\quad
 \\
  2(\r'^2+\r^2 \p'^2)&=2\psib'\psi'\,,\quad  
2(\dot\r^2+\r^2 \dot\p^2)=2\psibd\psid\,,\quad
2(\r'\dot\r+\r^2 \p'\dot\p) =\psib' \psid+\psibd \psi' \,. 
 \eal
 This is exactly what we have in \eqref{LTTbmnsc}, and the only question remaining is what happens with the potential $U$ in the limit. Expanding the potential in powers of $\r$, one gets
 \bal
 U = (J_3-J +\eta\nu)\r^2 -\frac{1}{6} \eta  (\eta\,  \nu -4 J+4 J_3)\rho ^4 + \cO(\r^6)\,.
 \eal
 Now, to reproduce the NLS model potential we impose the conditions
 \bal
 J_3-J +\eta\nu= -\mu\,,\quad -\frac{1}{6} \eta  (\eta\,  \nu -4 J+4 J_3)=\k\,,
 \eal
 and get
 \bal
 J_3=J+\frac{\mu}{3}-\frac{2 \k}{\eta }\,,\quad \nu = \frac{2 \k}{\eta ^2}-\frac{4\mu}{3 \eta }\,.
 \eal
It is then easy to check that in the limit $\eta\to 0$
\bal
U\to \k \r^4-\mu \r^2=\k\, (\psib\psi)^2-\mu\, \psib\psi\,,
\eal 
which is indeed the NLS model potential.

The sine-Gordon model is also a limiting case of  the LL model. To get the SG model we 
set $U_{\rm add}=0$, and
 parametrise $S_k$ as   \cite{faddeev}
\bal
S_1=-{\b \pi\ov2}\,,\quad S_2 = \sqrt{R^2-{\b^2 \pi^2\ov4}} \sin {\b \p\ov2}\,,\quad S_3 =  \sqrt{R^2-{\b^2 \pi^2\ov4}} \cos {\b \p\ov2}\,,
\eal
where $\b$ is a new constant, and $\pi$ and $\p$ are the fields parametrising $S_k$. 
We then get
\bal
\rK^t_t &= P_k\dot S_k= \frac{\beta ^2 }{4\eta} \pi\dot\phi \,,\quad
 \rK^t_x = P_k S'_k= \frac{\beta ^2 }{4\eta} \pi\,\p'\,,\\
  {1\ov\eta R^2} S_k'^2&={1\ov\eta R^2}\frac{\beta ^2 \left(\phi'^2 \left(\beta ^2 \pi^2-4 R^2\right)^2+16 R^2 \pi'^2\right)}{64 R^2-16 \beta ^2 \pi^2}\,,
  \\  
  {1\ov\eta R^2} \dot S_k^2&={1\ov\eta R^2}\frac{\beta ^2 \left(\dot\phi^2 \left(\beta ^2 \pi^2-4 R^2\right)^2+16 R^2 \dot\pi^2\right)}{64 R^2-16 \beta ^2 \pi^2}\,,
  \\
  {1\ov\eta R^2} S_k'\dot S_k&={1\ov\eta R^2} \frac{\beta ^2 \left(\phi'\dot\phi \left(\beta ^2 \pi^2-4
   R^2\right)^2+16 R^2 \pi'\dot\pi\right)}{64 R^2-16 \beta ^2 \pi^2}\,,\quad 
   \\
  U=-&\frac{\beta ^2 \pi^2 \left(J_2 (\cos (\beta  \phi)-1)-J_3 (\cos (\beta  \phi)+1)+2
   J_1\right)+4 \left(J_3-J_2\right) R^2 (\cos (\beta  \phi)-1)}{16 \eta }\,.
\eal
Now, we choose $\eta = \frac{\beta ^2 }{4} $,  take the limit $R\to\infty$, and get
\bal
\rK^t_t &\to \pi\dot\phi \,,\quad
 \rK^t_x \to\pi\,\p'\,,\quad
  {1\ov\eta R^2} S_k'^2\to\phi'^2\,,
  \quad
  {1\ov\eta R^2} \dot S_k^2\to\dot\phi^2\,,
  \quad
  {1\ov\eta R^2} S_k'\dot S_k\to\phi'\dot\phi \,. 
   \eal
   Finally, we choose $J_k$ as \cite{faddeev}
   \bal
   J_2=J_1+1\,,\quad J_3=J_1+1+{m^2\ov R^2}\,,
   \eal
   and in the limit $R\to\infty$ get $U$
   \bal
  U&=\frac{1}{2} \pi^2+\frac{m^2}{\beta ^2} (1-\cos \beta  \phi )\,.
\eal
Thus, in this limit we get the \TTb\ deformation of a model with the seed Lagrangian 
\bal
\cL_0 = \pi\dot\p -{1\ov2}\p'^2 - \frac{1}{2} \pi^2-\frac{m^2}{\beta ^2} (1-\cos \beta  \phi )\,,
\eal
which is indeed the SG model Lagrangian.
%%%%%%%%%%%%%%%%%%%%%%%%%%%%%%%%%%%%%%%%%%
\subsection{\texorpdfstring\TTb\ \ deformed Gardner equation}\la{gardner}
%%%%%%%%%%%%%%%%%%%%%%%%%%%%%%%%%%%%%%%%%%
The Gardner equation is a  combined KdV-mKdV equation
\bal
\dot u+ \mu\, u'  + 6\,g\, u u' - 6\, h\, u^2u' + u'''=0\,,
\eal
where $g,h$ and $\mu$ are  constants. If $u$ satisfies periodic boundary conditions then  $\mu$ can be removed by a constant shift of $u$
\bal
u\to u - c\,,\quad h c^2 + g c - {\mu\ov6}=0\,,
\eal
which also changes $g$. For decreasing boundary conditions such a shift is obviously forbidden. The Gardner equation is the continuity equation for the current
\bal\la{kdvcurrent}
J^t=u\,,\quad J^x = \mu\, u  + 3\,g\, u^2 - 2\, h\, u^3 + u''\,,
\eal
and if the charge $Q=\int dx \,u$ exists then it is conserved. In what follows we only consider the case where $Q$ exists.

The  Gardner equation can be derived from the action
\bal\la{gard}
S_0 = \int\, {\rm
d}x{\rm d}t\, \cL_0\,,\quad \cL_0=\k\,(-\dot\p\p' -\mu \p'^2 -2g\p'^3+ {h}\p'^4+\p''^2\,)\,,
\eal
where the field $\p$ satisfies the boundary conditions 
\bal\la{phibc}
\p(t,\infty) - \p(t,-\infty) =Q_\p=\const\,,
\eal 
$\k$ is any constant, and $u$ is related to $\p$ as
\bal
u=\p'\,.
\eal
Obviously, in the undeformed case $Q_\p =Q$. The equation of motion for $\p$ is invariant under a shift of $\p$ by any function of time. By using this invariance one may require $\p(t,\pm\infty)$ to be constant. However, as we will see, in the deformed case this invariance is broken, and different time dependence of $\p(t,\infty)$ leads to different solutions.

\medskip

To write the Lagrangian \eqref{gard} in the form \eqref{Sseed}, we first introduce an auxiliary field $A$ satisfying the equation of motion $A=\p''$, and cast $\cL_0$ into the form
\bal\la{gard2}
\cL_0&=\k\,(-\dot\p\p' -\mu \p'^2 -2g\p'^3+ {h}\p'^4+2A\p'' - A^2)
\,.
\eal
Then, we introduce  auxiliary fields for $\p'$ and $\dot\p$
\bal\la{gard3}
\un=-{1\ov\k}{\pa \cL_0\ov \pa\dot\p}=\p' \,,\quad B=-{1\ov\k}{\pa \cL_0\ov \pa\p'}=\dot\p +2\mu \p'+6g\p'^2- 4h\p'^3+2A'\,,
\eal
and get the desired form of the Lagrangian
\bal\la{gard5}
\cL_0 =\k\,(-\un\,\dot\p-B\p'+2A\,\un' +\un\,B -\mu\,\un^2-2g\,\un^3+ h\,\un^4 -A^2 )\,.
\eal
Clearly, the auxiliary field $\un$ is the Gardner field $u$, and the existence of the conserved current \eqref{kdvcurrent} is the consequence of the invariance of $\cL_0$ under constant shifts of $\p$.

We see that the set $(\Psi^a)$ consists of $\p,\un,B,A$, and 
\bal
\rK^t_t &= -\k \, \un\,\dot\p\,,\quad \rK^x_x =-\k \, B\p'  +2\k\,A\,\un'\,,\quad
 \rK^t_x =-\k\,\un\,\p'\,,\quad \rK^x_t = -\k \,B\dot\p  +2\k \,A\,\dot \un\,, 
\quad
\\
V &=-\k\,(\un\,B -\mu\,\un^2-2g\,\un^3+ h\,\un^4 -A^2 )\,.
\eal
Therefore, the \TTb\ deformed Lagrangian of the Gardner model is
\bal\la{Lttbgard}
\cL=\k {-\un\,\dot\p-B\p'+2A\,\un' +\un\,B -\mu\,\un^2-2g\,\un^3+ h\,\un^4 -A^2  -2\a\k\, A\,\un\,(\un'\dot\p -\dot \un\,\p') \ov 1-\a\k\,(\un\,B -\mu\,\un^2-2g\,\un^3+ h\,\un^4 -A^2 )}\,,
\eal 
where the field $\p$ satisfies the same boundary conditions \eqref{phibc} as in the undeformed case.  The undeformed Lagrangian \eqref{gard5} changes under the transformation
\bal
\p\to\p+f(t)\,,\quad B\to B + {df\ov dt}\,,
\eal
by a derivative term
\bal
\cL_0\to\cL_0 - \k {\pa\ov \pa x}\big(\,{df\ov dt}\p\,\big)\,.
\eal
The \TTb\ deformed Lagrangian \eqref{Lttbgard}, however, transforms in a nontrivial way, and therefore the time dependence of $\p$ at $x=\pm\infty$ changes physical properties of the \TTb\ deformed Gardner model.

In the undeformed model the auxiliary field $\un$ coincides with the Gardner field $u$. 
It is therefore reasonable to  use the same identification in the \TTb\  deformed Lagrangian \eqref{gard5}. 
One might try to use the fact that the Gardner equation is the continuity equation, and to
identify $\p'$ or  $\cJ^t=-{1\ov\k}{\pa \cL\ov \pa\dot\p}$ with $u$. Both $\p'$ and $\cJ^t$ are time components of conserved currents and coincide with $u$ in the undeformed case. 
Our analysis of the one-soliton solution of the \TTb\ deformed KdV equation  indicates that the auxiliary field $\un$ is a better choice. 

It is impossible to get rid of all the auxiliary fields and get a local Lagrangian because the Lagrangian depends on derivatives of $\un$. In what follows without loss of generality we set $\k=1$.

%%%%%%%%%%%%%%%%%%%%
\subsection{Comments}\la{comm1}
%%%%%%%%%%%%%%%%%%%%%

Here we discuss  similarities and differences of the \TTb\ deformed Lagrangians for relativistic and non-relativistic models obtained in this section, and comment on possible approaches to quantising the models.

\smallskip

All the Lagrangians depend on auxiliary fields which are introduced in a seed model to bring it to the first-order form \eqref{Sseed}. If the physical fields of a seed model do not depend on second- or higher-order  derivatives then  auxiliary fields enter a \TTb\ deformed Lagrangian algebraically, and can be eliminated leading in the cases considered to Nambu-Goto type actions.  More complicated seed models (even relativistic invariant) may lead to \TTb\ deformed Lagrangians which are solutions to high degree polynomial equations. 

\smallskip

A Nambu-Goto type Lagrangian obtained by eliminating auxiliary fields has a square root sign ambiguity. If a model is considered on a line then the requirement of finiteness of the energy singles out the perturbative in $\a$ branch of the deformed Lagrangian depending only on the physical fields. However, if the model is on a circle then one has to find additional requirements to single out the perturbative  branch.  
For example if one considers the \TTb\ deformed free massless scalars and chooses the negative sign in front of the square root in the \TTb\ deformed Lagrangian \eqref{LbosTTb} then  for  $\a<0$ the energy is not bounded from below. In quantum theory it would clearly be unsatisfactory. On the other hand if $\a>0$ then  the energy of any solution is bounded from below,\footnote{Note that if there is no potential and  $B$-field then the equations of motion do not depend on the branch of the square root. } and diverges in the limit $\a\to0$. Thus, if one calculates, for example, the partition function of the \TTb\ deformed model then there seems to be no reason not to include the contribution from the nonperturbative branch to the path integral over physical and auxiliary fields. It would imply that for $\a>0$ the spectrum of \TTb\ deformed relativistic models previously discussed is incomplete and must be supplemented by a nonperturbative part. 

\smallskip

The physical fields of the Gardner model depend on second-order  derivatives. As a result the \TTb\ deformed  equations of motion for the auxiliary fields are not algebraic, and depend on space derivatives of the auxiliary fields. Eliminating the auxiliary fields (which we have not managed to do) would lead to an action non-local in space.  The \TTb\ deformed Gardner model is, therefore, expected to have properties noticeably different from the seed model already at the classical level. Indeed in the next section we will see that solutions of the \TTb\ deformed KdV equation are very sensitive to the behaviour of the field $\p$ at space infinities.

\smallskip

 We have seen that the deformation drastically modifies the Poisson structure of all the non-relativistic models we considered, and developing a Hamiltonian formulation requires dealing with an intricate system of second-class constraints. This actually makes \TTb\ deformed non-relativistic models more complicated than the relativistic ones where the Hamiltonian formulation is straightforward. 
 
 \smallskip
 
 One may wonder whether the \TTb\ non-relativistic deformed models exist as quantum theories.  We do not expect any principal difficulties in perturbative quantisation of the \TTb\ deformed NLS and LL models. For example, the expansion of the deformed Lagrangian  \eqref{LTTbmnsc} of the NLS model in powers of $\a$ is straightforward, and the standard technique can be used to compute the scattering matrix. It is expected that the \TTb\ deformed S-matrix would be different from the undeformed one only by the \TTb\ CDD  factor.  It might be necessary to tune properly counterterms 
but the relation between the S-matrices is very general and should be considered as a part of the definition of a quantised \TTb\ deformed model.
For integrable models the relation follows from the interpretation, discussed at length in \cite{SF19a}, of the homogeneous inviscid Burgers equation as the condition of the gauge invariance of the target space-time energy and momentum of the string theory which produces a \TTb\ deformed model in $\a$-dependent light-cone gauge. The UV behaviour of the \TTb\ deformed NLS model should be milder than for relativistic ones because of the absence of virtual particles production.
 
The spectrum of the \TTb\ deformed NLS (and LL) model on a circle can be also studied perturbatively. 
 At each order in $\a$ one can remove all interaction terms with time derivatives of $\psi$ by a field redefinition producing new terms with higher space derivatives.    The resulting model has the undeformed Poisson structure and can be easily quantised. The spectrum of the Hamiltonian can then be found as an expansion in powers of $\a$. 
 
\smallskip 

For finite $\a$
 another, more pragmatic, approach to the \TTb\ deformed spectrum is to postulate that it is governed by the usual Bethe equations with the \TTb\ deformed S-matrix. It was done in \cite{Jiang2020a} for the deformed NLS model in the repulsive regime, and it was found that the properties of the model were similar to the properties of  \TTb\ deformed CFT's. In particular, 
 for $\a<0$ the spectrum is well-defined but there exists an upper bound for the temperature while for $\a>0$ there exists a critical value $\a_{\rm c}$ which depends on the number of particles and the radius such that for $\a>\a_{\rm c}$ the spectrum becomes complex.  
 However, there is no argumentation why the Bethe equations would not be replaced by a more complicated system of TBA-like equations.  It would be interesting to compute the spectrum as an expansion in powers of $\a$, and compared it with the Bethe ansatz predictions.  
 %Clearly, studying quantum aspects of \TTb\ deformed non-relativistic models is beyond the scope of our paper.

\smallskip

To conclude this subsection let us mention that 
we do not think that  the approaches discussed above can  be applied to quantum \TTb\ deformed Gardner model.  
In the simplest KdV case the spectrum of quantum KdV theory is described by massless TBA equations \cite{BLZ94}  which are derived by quantising the second Hamiltonian structure of the KdV equation and diagonalising the infinite-dimensional abelian  subalgebra of the Virasoro algebra which gives commuting integrals of motion of quantum KdV model. It is unclear how these deep relations are modified under the \TTb\ deformation, and we suspect that quantum (and even classical) \TTb\ deformed KdV model may hide many surprises.

%%%%%%%%%%%%%%%%%%%%%%%%%%%%%%%%%%%%%%%%%%
\section{Deformed one-soliton solutions}

In this section we derive
one-soliton solutions of the deformed NLS and KdV  models in order to see whether they exhibits the general phenomenon of widening/narrowing the width of particles under the \TTb\ deformation recently discussed in \cite{Cardy2020}. 

%%%%%%%%%%%%%%%%%%%%
\subsection{\texorpdfstring\TTb\  \ deformed NLS soliton}
%%%%%%%%%%%%%%%%%%%%%

In this subsection we discuss a one-soliton solution of the \TTb\ deformed NLS model. Let us first  recall some properties of the  seed  model. Its Lagrangian  is given by \eqref{LmNS2}  where  $\psi,\psib$ (and $A,\Ab$) are complex fields  conjugate  to each other. The Lagrangian is invariant under the Galilean transformations
\bal
x\to x - v\,t\,,\quad t\to t\,,\quad \psi\to e^{ \frac{i}{4} v^2t-\frac{i}{2}v\,x }\psi\,,\quad A\to  e^{\frac{i}{4} v^2t-\frac{i}{2}v\,x }(A-\frac{i}{2}v\,\psi)
\eal
which implies the usual nonrelativistic dispersion relation for a one-soliton solution, and allows one to recover a full solution from a soliton at rest.  It is also invariant under the U(1) transformations $\psi\to e^{i\z}\psi\,,\, A\to e^{i\z}A$,  and the finite density term $\mu\psib\psi$ is proportional to the time component of the conserved U(1) current.  It can therefore be removed by the following time-dependent U(1) transformation
\bal\la{u1q}
\psi\to e^{-i\mu\, t}\psi\,,\quad A\to e^{-i\mu\, t}A\,.
\eal
Thus, in the rapidly decreasing case the finite density term plays no essential role in the undeformed NLS model. 

The one-soliton solution we are going to deform exists for $\k<0$, and to simplify the formulae below we introduce a new coupling constant $g>0$ related to $\k$ as 
\bal
\k = -{g^2\ov4}\,.
\eal  
Then, the one-soliton solution  is given by 
\bal\la{soliton0}
\psi = 
 {u\ov g}\, {1\ov \cosh\big({u\ov2}(x-vt)\big)}e^{i\p}\,,\quad \p & = \frac{v}{2} (x-vt) + \frac{t}{4}  \left(u^2+v^2+4\mu\right) \,,\quad A=\psi'\,,
\eal
where $v$ is the velocity of the soliton, and $u>0$ can be chosen to be positive without loss of generality.

The U(1) charge $Q$, the momentum $P$ and the energy $E$ of the soliton are 
\bal\la{charges}
Q&=\int_{-\infty}^\infty dx\,\psib\psi = {4u\ov g^2}\,,
\\
P&=-\int_{-\infty}^\infty dx\, T^t{}_x =\frac{2u \,v}{g^2 } = m\,v\,,\quad m=\frac{2u}{g^2 }=\frac{Q}{2 }\,,
\\
E & =\int_{-\infty}^\infty dx\, T^t{}_t =\frac{u v^2}{g^2 }-\frac{u^3}{3g^2 }- \frac{4u\mu}{g^2 }= \frac{ P^2}{2m}-\frac{1}{24} g^4 m^3-\mu\,Q\,,
\eal
and up to a constant the dispersion relation is indeed  nonrelativistic,  and the U(1) charge is twice the mass of the soliton.

\medskip

To find a \TTb\ deformation of the soliton \eqref{soliton0}, we begin with 
the \TTb\ deformed Lagrangian 
\eqref{Lttbnmsc} which for  the NLS  model  simplifies to
\bal\la{LNLS}
\cL= \frac{\frac{i}{2}(\psib \psid - \psibd \psi) - \bar{A}\psi' -  \psib' A+ \bar{A}A - U + \alpha\frac{i}{2}(\bar{A}\psi +  \psib A)(\psibd \psi' - \psib'\psid)}{1-\alpha(\bar{A}A - U)}\,,
\eal
\bal
U=-{g^2\ov4}(\psib \psi)^2 - \mu\psib \psi\,.
\eal
It is clear from the Lagrangian \eqref{LNLS} that the U(1) transformation \eqref{u1q} does not remove the $\mu$-dependent terms, and therefore,  \TTb\ deformed soliton properties  depend  on it.

It is convenient to introduce the polar coordinates for $\psi$ and redefine the auxiliary fields as follows\footnote{These variables are also useful for analysing the $JT$-type deformations  \cite{Guica17} of the NLS model.}
\bal
\psi = \rho\, e^{i \phi}, \quad \psib = \rho\, e^{-i \phi}, \quad A = \rho_A e^{i \phi} \,,\quad \Ab = \bar\rho_A e^{-i \phi}\,,
\eal
because the U(1) symmetry is realised just by shifts of $\p$, and the Lagrangian depends only on the derivatives of $\p$. Clearly, $\r$ is the amplitude and $\p$ is the phase of the soliton.
In terms of the fields the Lagrangian \eqref{LNLS} takes the form 
\bal\la{LNLS2}
\cL= \frac{-\r^2\dot\p - (\rho_A+\bar\rho_A)\rhop + i(\rho_A-\bar\rho_A)  \rho\, \phip  +\bar\rho_A\rho_A +{g^2\ov4} \r^4 + \mu\, \r^2 -\a \rho ^2 (\rho_A+\bar\rho_A)(\rhod\phip-\rhop\phid)}{1-\alpha(\bar\rho_A\rho_A +{g^2\ov4} \r^4 + \mu\, \r^2)}\,,
\eal
where  $\r_A$ and $\bar\r_A$ are complex conjugate to each other.

\medskip

The deformed one-soliton solution can be derived by explicitly solving the equations of motion by using the following ansatz
\bal
\r(t, x) & = \rho(x-vt) \,, \quad \r_A(t, x) = \rho_A(x-vt) \,,\quad  \rho(\pm \infty) = 0\,,\\
\p & =\w\,t  + \vp(x-vt)\,,\quad \w=\frac{u^2+v^2}{4} +\mu\,.
\label{eq:soliton_ansatz}
\eal
 The phase $\p$ of the soliton is at most the sum of a linear function of $x, t$ which we can choose without loss of generality to be the same as in the undeformed case,  and of a function of $x-vt$ due to the restricted dependence of the other fields. 
  
\medskip

The derivation  is sketched in appendix \ref{NLSsol}, and the solution can be expressed in terms of $\rho$ as follows
\bal
\rho' & = \pm \frac{2 \rho \sqrt{u^2-g^2 \rho^2}}{4+\a  \rho^2 \left(-2g^2 \rho^2+u^2-v^2-4\mu\right)}, \quad \rho _A = \frac{1}{2} \rho \left(i v \pm \sqrt{u^2-g^2 \rho^2}\right)\,,\\
x-vt & = x_0 \pm \frac{2 \coth ^{-1}\left(\frac{u}{\sqrt{u^2-g^2 \rho^2}}\right)}{u} \mp \frac{\a  \sqrt{u^2-g^2 \rho^2} \left(u^2+3 v^2+12\mu+2g^2 \rho^2\right)}{6g^2 }\,,
\\
\p & = \frac{1}{2} v (x-vt) + \frac{1}{4} t \left(u^2+v^2+4\mu\right) \pm \frac{\a  v \left(u^2-g^2 \rho^2\right)^{3/2}}{6g^2}\,.
\label{eq:soliton_sol}
\eal
Since the phase $\p$ and the auxiliary field $\r_A$ are smooth functions of $x$ and $t$ if the amplitude $\r$ is, we discuss only the properties of $\r$. Unlike the undeformed soliton, the amplitude has a nontrivial dependence on the chemical potential $\mu$.  However, it enters the amplitude only  through the combination $v^2 + 4\mu$. 
Without 
 loss of generality we can set $t=0$ and $x_0=0$. Clearly,  the maximum of $\r(x)$ is equal to $u/g$, and it is at $x=0$. 
 From the equation for $\r'$ we see that $\r$ is a single-valued function of $x$ only if $\r'\neq\infty$ for all $x$ which leads to the condition
 \bal\la{rpinfty}
 4+\a  \rho^2 \left(-2g^2 \rho^2+u^2-v^2-4\mu\right)\neq  0\quad \text{for} \quad 0\le\r\le{u\ov g}\,.
 \eal
To analyse \eqref{rpinfty} it is convenient to introduce a new parameter
\bal
W={u^2-v^2-4 \mu} \,.
\eal
Then, the roots of the equation $\r'=\infty$ are given by
\bal
\r^2_\pm = \frac{W\pm \sqrt{\frac{32 g^2}{\alpha }+W^2}}{4g^2}\,,\quad \r'\big|_{\r=\r_\pm}=\infty\,.
\eal
A simple analysis shows that the roots $\r_\pm$ are outside the interval $(0\,,\, u/g)$ if
\bal
\text{I.}&\quad W\in \bR\quad \text{and}\quad  -{32g^2\ov W^2}<\a\le 0 \quad\Rightarrow\quad \text{complex}\ \ \r_\pm^2
\\
\text{II.}&\quad W=u^2-v^2-4 \mu< 0\quad \text{and}\quad \a\le  -{32g^2\ov W^2} \quad\Rightarrow\quad \r_-^2\le\r_+^2<0
\\
\text{III.}&\quad W-4u^2=-3u^2-v^2-4 \mu > 0\quad \text{and}\quad 
\\
&\quad\frac{4 g^2}{2 u^4-u^2 W}<\alpha \le -\frac{32 g^2}{W^2}\quad\Rightarrow\quad {u^2\ov g^2}<\r_-^2\le\r_+^2
\\
\text{IV.}&\quad W-2u^2=-u^2-v^2-4 \mu < 0\quad \text{and}\quad 
\\
&\quad 0<\alpha <\frac{4 g^2}{2 u^4-u^2 W} \quad\Rightarrow\quad \r_-^2<0<{u^2\ov g^2}<\r_+^2
\\
\text{V.}&\quad W-2u^2=-u^2-v^2-4 \mu > 0\quad \text{and}\quad 
 \alpha >0 \quad\Rightarrow\quad \r_-^2<0<{u^2\ov g^2}<\r_+^2
\eal
Introducing the following two critical values of $\a$
\bal\la{apm}
\a_-\equiv -{32g^2\ov (u^2-v^2-4 \mu)^2}<0\,,\quad  \a_+\equiv \frac{4 g^2}{u^2(u^2+v^2+4 \mu)} \,,
\eal
we can combine these regions as follows
\bal\la{ABC}
\text{A.}&\quad  -\infty<u^2-v^2-4 \mu < 0\quad \text{and}\quad 
 -\infty<\alpha <\a_+\,,\quad \a_+ >0
 \\
\text{B.}&\quad\ \ \,  0<u^2-v^2-4 \mu <2u^2\quad \text{and}\quad 
\a_-< \alpha <\a_+\,,\quad \a_+ >0
 \\
\text{C.}&\quad 2u^2<u^2-v^2-4 \mu <4u^2\quad \text{and}\quad 
 \a_+<\a_-<\alpha <\infty
\\
\text{D.}&\quad 4u^2<u^2-v^2-4 \mu <\infty\quad \text{and}\quad \a_+<\alpha  <\infty\,,\quad \a_+<\a_-<0
\eal
The condition A is satisfied if $v^2>u^2-4\mu$ which imposes a lower bound on $v^2$ if  $u^2>4\mu$.
If $\mu\ge0$ then the condition B is satisfied for all $u\,,\,v$ but C and D are never satisfied.
The condition D is satisfied if $v^2<-3u^2-4\mu$ which imposes an upper bound on $v^2$ if  $3u^2<-4\mu$.
 If $\m<0$ then all the four conditions can occur.
 
\begin{figure}
    \centering
    \includegraphics[width = 0.32\linewidth]{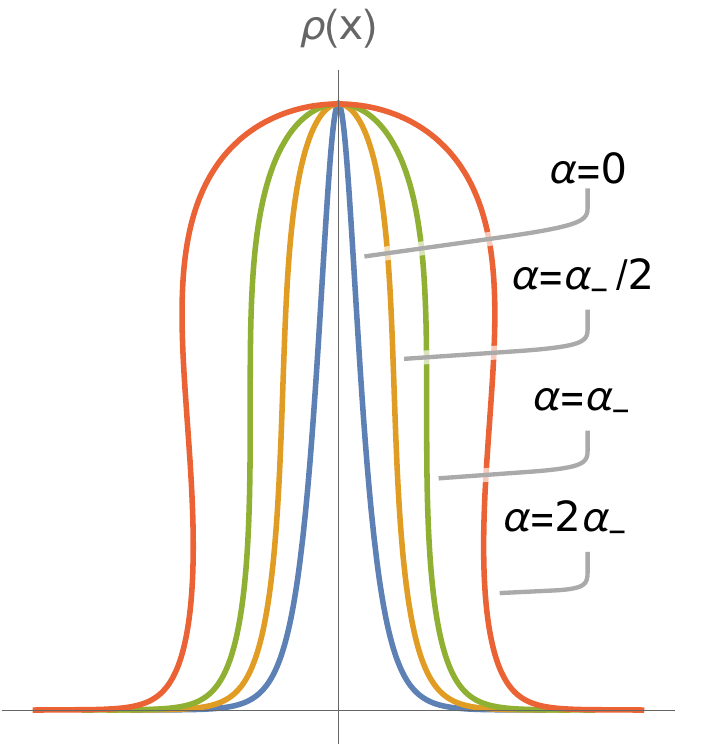}
    \includegraphics[width = 0.32\linewidth]{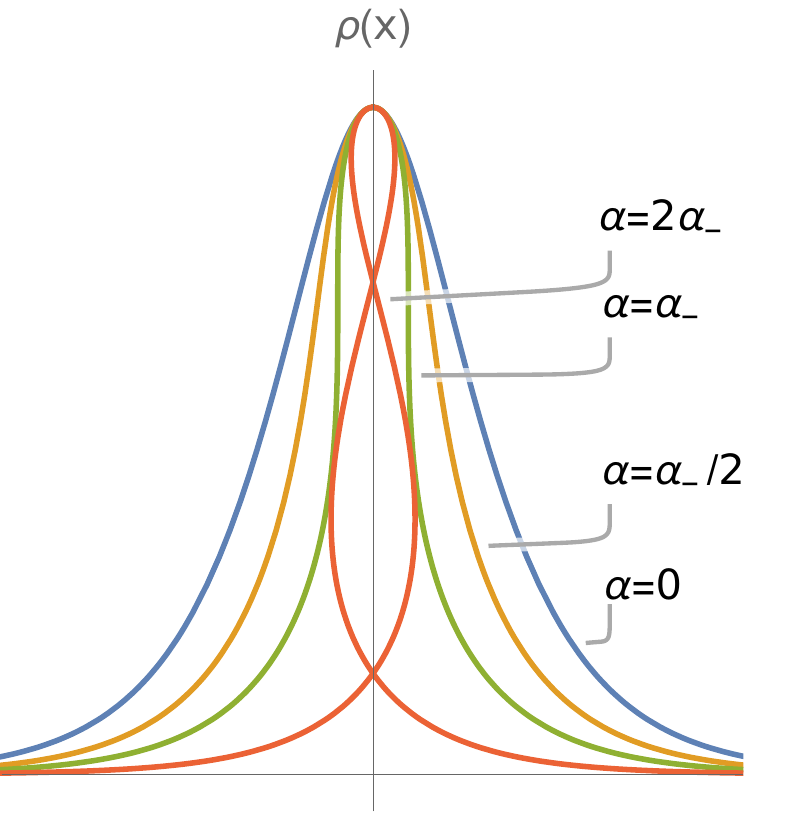}
    \includegraphics[width = 0.32\linewidth]{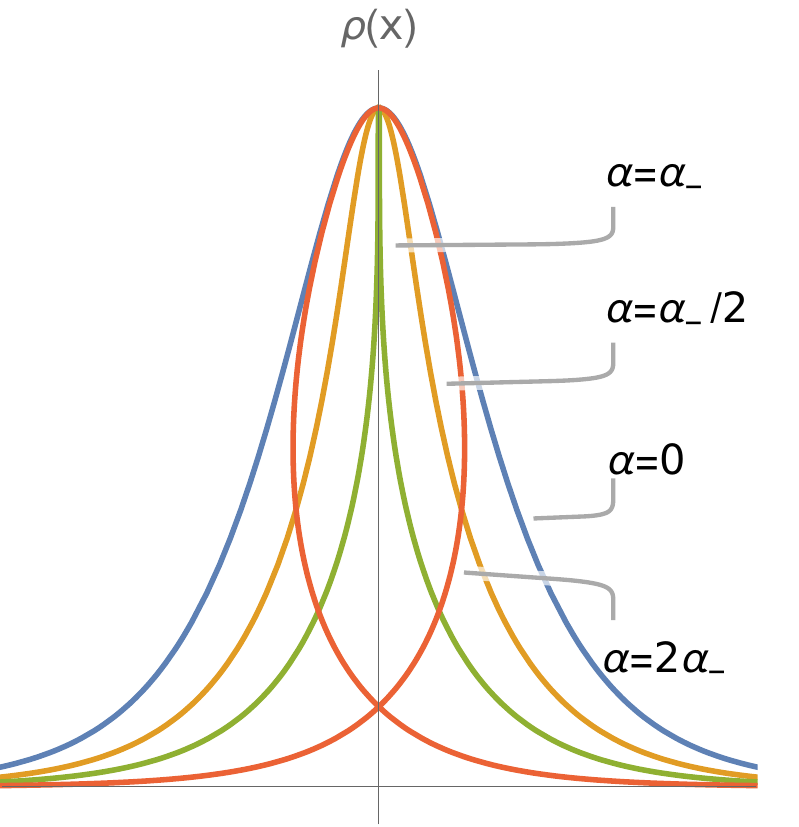}
     \caption{Left: Case B, $\m = 0$, $\a_- = -32$, displaying formation of shockwave solution for negative $\a$. Centre: Boundary case of B and C, $\m = -1/4$, $\a_- = -8$, example of competing shockwave and narrowing behaviours creating a double-loop solution. Right: Case C, $\m = -0.6$, $\a_- = -2.76817$, soliton is becoming singular at $\a_-$, after which it forms a loop. }
    \label{fig:NLSMain}
\end{figure}

If the parameters of the soliton satisfy one of the conditions \eqref{ABC} then $\r(x)$ is an even function of $x$, and 
the differential equation for $\rho$ allows one to replace the integration of any expression over $x$ with the integration over $\rho$. The U(1) charge,  energy and  momentum of the soliton are  easily found, appear to be unchanged by the deformation, and are given by \eqref{charges}. The shape of the soliton obviously changes, and, in particular, we can  define its size by using the full-width-half-maximum
\bal\la{fwhm}
FWHM & =- \alpha\frac{\sqrt{3} \,  u \left(u^2+2 v^2+8\mu\right)}{4g^2}+\frac{4 \log \left(2+\sqrt{3}\right)}{u}\,.
\eal
The soliton clearly exhibits the general phenomenon of widening/narrowing the width of particles under the \TTb\ deformation \cite{Cardy2020}. However, whether the size is increasing or decreasing depends not only on the sign of $\a$ but also on the sign of $s\equiv u^2+2 v^2+8\mu$. Obviously,  it is positive for all values of $u$ and $v$ only if $\mu \ge 0$. It is also positive if the soliton parameters satisfy condition A but  it is negative for conditions C or D.  For parameters satisfying condition B one can have both positive and negative $s$ if $\mu$ is negative.
The visually distinct solutions are demonstrated in figure \ref{fig:NLSMain}. Further plots for all cases are shown in figures \ref{fig:NLSA} and \ref{fig:NLSD} in Appendix \ref{NLSfig}, which display the same behaviours as in case C. Since the amplitude depends only on $v^2 + 4\mu$ we set $v=0$ without loss of generality when plotting solutions. We set $g=1, \ u = 1$, so that the graphs are parametrised by $\m$.
If the soliton base widens (or remains constant if $u^2+2 v^2+8\mu=0$)  as the magnitude of $\a$ increases then the peak flattens as in figures \ref{fig:NLSA} and \ref{fig:NLSD}. 
 Let us also mention that as one can see from \eqref{fwhm} the heavier and speedier the soliton is the wider it is. That is very different from the undeformed case where the width is independent of  speed and decreases with mass increasing.

Let us now assume that $u,v,\mu$ satisfy one of the conditions \eqref{ABC} but $\a$ is at a boundary of its allowed values, i.e. it takes one of the critical values $\a_\pm$. Then, a shock-wave singularity develops, and  away from the critical values the solution $\r(x)$ becomes a multi-valued function of $x$. In this case at least one of the roots $\r_\pm$ is inside the interval $(0\,,\, u/g)$. Regions where only one root exists form loops as in figures \ref{fig:NLSMain}, \ref{fig:NLSD} and \ref{fig:NLSA}, due to $x(\rho)$ (given explicitly in \eqref{eq:soliton_sol2}) becoming negative. This happens if $u,v,\mu$ satisfy either conditions A and B with $\a>\a_+>0$ or C and D with $\a<\a_+<0$. Where both roots exists the solution is either a bell shape or a double loop shape, both shown in figure \ref{fig:NLSMain}. This happens if $u,v,\mu$ satisfy condition B with $\a<\a_-<0$, or condition C with $\a_+<\a<\a_-<0$. The conditions for the appearance of these solutions are summarised below
\bal
\text{Loop: }&
\left\{
\begin{array}{cc}
 u^2+v^2+4 \mu >0\,, \ \a > \a_+ >0\,, & \ \r_-<0<\r_+<{u\ov g}    \\
 u^2+v^2+4 \mu <0\,, \ \a < \a_+<0\,, &\  0<\r_-<{u\ov g} <\r_+   
\end{array}
\right.
\\
\text{Bell or Double Loop: }&
\left\{
\begin{array}{cc}
   0<u^2+v^2+4 \mu <2u^2\,,& \ \a < \a_-<0  <\a_+  \\
-2u^2<u^2+v^2+4 \mu <0\,, &\ \a_+ < \a < \a_-<0  
\end{array}
\right.
\eal

The existence of the forbidden regions of the soliton parameters is disturbing because the soliton physical quantities such as mass, momentum and energy do not show any sign of singularity. It is also unclear if it is compatible with
 the integrability preserving feature of the \TTb\ deformation. We attempt to fix this by redefining the amplitude function as a piecewise smooth curve by exploiting the translational invariance  of $x-vt$.

\begin{figure}[h!]
    \centering
    \includegraphics[width = 0.3\linewidth]{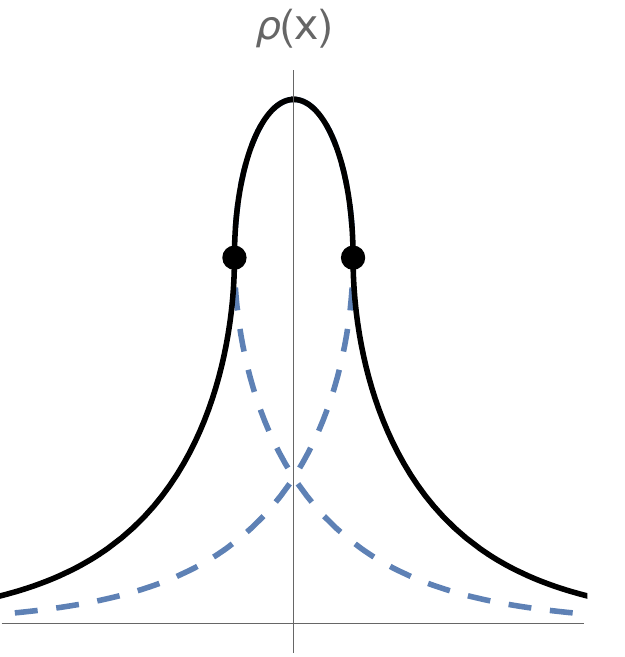}
    \includegraphics[width = 0.3\linewidth]{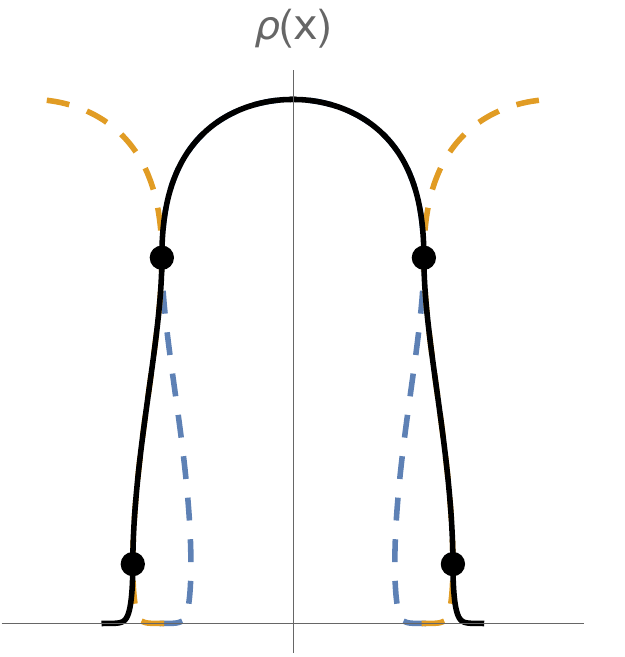}
     \includegraphics[width = 0.3\linewidth]{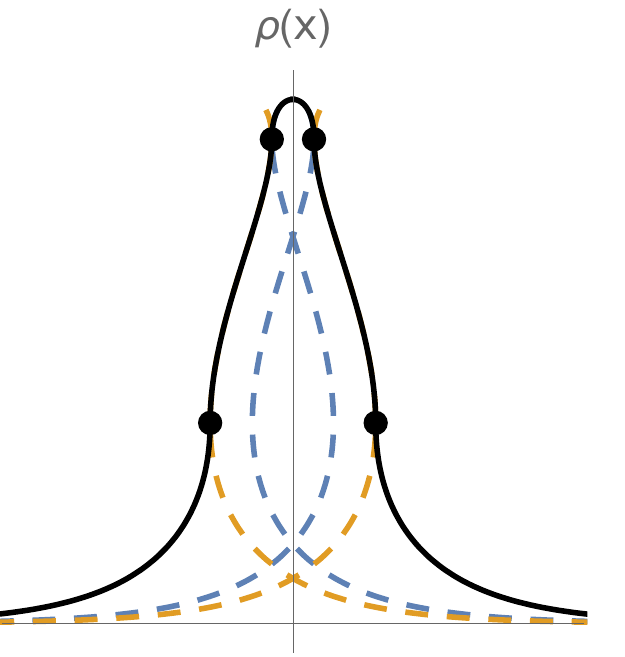}
    \caption{Demonstration of the gluing procedure on the loop (Left), bell (Centre) and double-loop (Right) soliton solutions, indicating the points where $\r'$ becomes singular. }
    \label{fig:NLSGlue}
\end{figure}
 
We set $t=0$, $x_0=0$, choose the upper sign in the solution \eqref{eq:soliton_sol} corresponding to the positive branch of the undeformed solution, and introduce the function
\bal
x(\r)& = \frac{2 \coth ^{-1}\left(\frac{u}{\sqrt{u^2-g^2 \rho^2}}\right)}{u} - \frac{\a  \sqrt{u^2-g^2 \rho^2} \left(u^2+3 v^2+12\mu+2g^2 \rho^2\right)}{6g^2 }\,,\quad 0\le\r\le {u\ov g}\,.
\label{eq:soliton_sol2}
\eal
 In terms of $x(\r)$ the piece-wise smooth solutions can be written as 
 \bal\la{forbidden_sol}
\text{Loop: }&
\left\{
\begin{array}{c}
x_L^+(\r)= -x(\r)\theta(\r-\r_+)+\big(x(\r)-2x_+\big)\theta(\r_+-\r)  \\
x_L^-(\r)= -x(\r)\theta(\r-\r_-)+\big(x(\r)-2x_-\big)\theta(\r_--\r)   
\end{array}
\right.
\\
\left.
\begin{array}{c}
\text{Bell:}  \\
\text{Double Loop:}  
\end{array}
\right\} &\ \  \left.
\begin{array}{c}
x_B(\r)= x(\r)\theta(\r-\r_+)+\big(2x_+-x(\r)\big)\theta(\r_+-\r) \theta(\r-\r_-)  \\
+\big(x(\r) + 2x_+-2x_-\big)\theta(\r_--\r)  
\end{array}
\right. 
\eal
where $\theta$ is the Heaviside function and $x_\pm = x(\r_\pm)$. 
Each of these functions is a positive decreasing function of $\r$ with a continuous first derivative. 
The soliton profile $\r(x)$ is an even function of $x$ given for $x\ge0$ by the functions inverse to \eqref{forbidden_sol}.
  The energy, momentum and charge densities are singular  as functions of $x$ but it is an integrable singularity. Since they depend on $\r'^2$ and $\r$, the energy, momentum and charge  are given by the same expressions \eqref{charges}. The three forbidden solution types are reconstructed into valid amplitudes in figure \ref{fig:NLSGlue}. Note that all these new solutions increase in width as $\a$ increases in magnitude. Whether such a gluing procedure is legitimate remains to be seen but there are examples of models with singular solitons, see e.g. \cite{pogr}.

\medskip

Let us finally mention that the inverse function $x(\rho)$ can also be derived through a dynamical coordinate transformation as described in \cite{Tateo18b}, and used in \cite{Tateo2020} to find the \TTb\ deformed one-soliton solution for the case $\mu=0$.

%%%%%%%%%%%%%%%%%%%%
\subsection{\texorpdfstring\TTb\  \ deformed KdV soliton}
%%%%%%%%%%%%%%%%%%%%%

In this subsection we discuss a one-soliton solution of the \TTb\ deformed KdV  equation which corresponds to the $g=1\,,\, h=0$ case of the Gardner equation
\bal
\dot u +\mu\, u'+ 6u u'+ u'''=0\,.
\eal
The constant $\mu$ is usually set to 0 but we prefer to keep it so that for $\mu<0$ we could have left-moving solitons.

The one-soliton solution we are going to deform is given by
\bal\la{solitonKdV}
u&={2 w^2\ov \cosh^2\big(w (x-vt)\big)}\,,\quad  w= \frac{1}{2}\sqrt{v-\mu} >0\,,
\\
\p&=2 w \tanh \left(w (x- vt)\right) +f(t)\,,
\eal
where $f(t)$ is any function of $t$.
As was discussed in the previous section, in the undeformed case the soliton properties are independent of  $f(t)$. In particular,
the charge $Q$, momentum $P$ and energy $E$ of the soliton are
\bal
Q&=\int_{-\infty}^\infty dx\, u = 4 w\,,\\
P&=\int_{-\infty}^\infty dx\, u^2 = {16\ov3} w^3 \,,\\
E&=\int_{-\infty}^\infty dx\, (\mu\, u^2+2u^3-u'^2) = \frac{16}{3} \mu w^3
+\frac{64}{5} w^5 = \mu  P+\frac{3}{5} \left(\frac{3}{2}\right)^{2/3} P^{5/3}\,.
\eal
A funny property of the soliton is that its momentum is always positive even if the velocity $v$ is negative which requires $\mu$ to be negative too. This is counter-intuitive and for $v<0$ it might be reasonable to change the overall sign of $P$ and $E$ which is equivalent to setting $\k=-1$ in the Lagrangian \eqref{gard5}. This also effectively changes the sign of $\a$ in the \TTb\ deformed Lagrangian \eqref{Lttbgard}.  Then, for small $P$ the dispersion relation would be approximately the one for a massless relativistic particle. In what follows to have a uniform description we will continue using $\k=1$ for all values of $v$.

\medskip

The \TTb\ deformed soliton solution  depends on the function $f(t)$ in a nontrivial way, and we only consider the simplest case $f(t) = b\,t$ where $b$ is an arbitrary constant. In fact, redefining $\p$ as $\p\to\p + bt$, we find that the \TTb\ deformed Lagrangian \eqref{Lttbgard} transforms as $\cL\to \cL - b\cJ^t$, and therefore $b$ can be interpreted as the parameter of the deformation by the time component of the conserved current due to the invariance of \eqref{Lttbgard} under  constant shifts of $\p$.

 In this case all auxiliary fields are only functions of $x-vt$, and the \TTb\ deformed solution can be found by using the equations of motion and the ansatz
\bal\la{KdV_ansatz}
\p = \p(x-v\,t) + b\,t\,,\quad \un = \un(x-v\,t)\,,\quad A = A(x-v\,t)\,,\quad B = B(x-v\,t)\,.
\eal
The full derivation is described in Appendix \ref{KdVsol}. We find that $\un$ rather than $\p'$ is the natural field to express our results in terms of. 
We define $\tw^2= {v-\mu - \a\,b^2\ov 4} = w^2 - {\a\,b^2\ov4}$ to simplify the following expressions.
%, yet this is not an arbitrary redefinition of the parameter $w$. 
The solution can be written as  a set of equations expressing $\un'$, $\p'$, $A$ and $B$ in terms of $\un$
\bal\la{unp}
\un' &=\pm \frac{ \un \sqrt{4\tw^2-2 \un}}{1+\a  \un^2 \left(4 \un-8\tw^2 - \a b^2\right)}\,,    & \quad  \p' &=\frac{\un-\a\, b \,\un^2}{1+\a  \un^2 \left(4 \un-8\tw^2 - \a b^2\right)} \,,\\
B &=\left(\m + 4\tw^2 \right) \un+b\,,  & \quad A &= \pm \un \sqrt{4\tw^2-2 \un}\,,\qquad \tw^2 = w^2 - {\a\,b^2\ov4}\,. \\
\eal
For the solutions to be real, $\tw^2 > 0$, or equivalently, $v>\mu +\a\, b^2$. For fixed $v,\mu,b$ this condition imposes an upper bound on allowed values of $\a$: ${\mu-v\ov  b^2}>\a$. Note also that for $\a<0$ one may have $w^2={v-\mu\ov4}<0$.

The extra parameter $b$  causes the deformed quantities of energy, momentum and the dispersion relation to be dependent on both $\a$ and $b$. 
\bal
E &= \frac{16}{15} \tw^3 \left(12\tw^2 + 5(\m -\a b^2)\right)\,, \quad  P = \frac{16}{3} \tw^3\,,\\
E(P) &= P \left(\m -\a  b^2\right)+\frac{3}{5} \left(\frac{3}{2}\right)^{2/3} P^{5/3}\,.
\eal
The appearance of $\a$  in the dispersion relation is due to the fact that the \TTb\ deformed KdV model is intrinsically nonlocal and sensitive to the boundary behaviour of $\p$.

Furthermore, the parameter $b$ causes the previously identical conserved charges of $\cJ^t$ and $\p'$ to become independent
\bal
Q &= \int dx \cJ^t = 4\tw \left(1+\frac{4}{3} \tw^2 \a b\right)\,, \qquad Q_\p = \int dx \p' = 4 \tw \left(1-\frac{4}{3} \tw^2 \a b\right)
\eal
We also find that $b$ defines a flow equation for a deformation under the current $\cJ^t$
\bal
\frac{\partial \cL}{\partial b} = -\frac{\un (\a  b \un+1)}{1-\a  \un^2 \left(-4 \un+\a  b^2+8 \tw^2\right)} = -\cJ^t\,.
\eal
Integrating the equation for $\un'$ in \eqref{unp}, we find the  inverse expression for $\un$
\bal\la{xun}
x-vt &= x_0 \,\pm\, \frac{\text{arctanh}\left(\frac{\sqrt{4 \tw^2-2\un}}{2 \tw}\right)}{\tw} 
\\
&\mp\frac{1}{15} \sqrt{2} \a  \sqrt{2 \tw^2-\un} \left(4 \left(2 \tw^2-\un\right) \left(3 \un+4 \tw^2\right)+5 \a  b^2 \left(\un+4 \tw^2\right)\right)\,,
\eal 
 which displays both shockwave and looping solutions as in the NLS case. 
 With $t=0$ and $x_0=0$ the maximum of $\un(x)$ occurs at $x=0$ for $\un(0) = 2 \tw^2$. 
 The full-width half-maximum of the soliton is
\bal
FWHM = \frac{2 \text{arcoth}\left(\sqrt{2}\right)}{\tw}-\frac{2\sqrt{2}}{15} \alpha  \tw^3 \left(25 \alpha  b^2+28 \tw^2\right)\,,
\eal
and for positive $\a$ it decreases. 

 \medskip
 
 The derivative $\un'$ becomes singular when the denominator in the equation for $\un'$ in \eqref{unp} vanishes 
\bal\la{eqcr}
d(\un)\equiv \alpha  \un^2 \left(4 \un-8 \tw^2\right. -\a  b^2)+1 = 0, \qquad 0 < \un < 2 \tw^2\,.
\eal
In much the same way as the NLS case, restricting the roots of this expression to lie inside the range of $\un$ will generate the conditions for the solution to become multi-valued.
 
A detailed analysis of the equation \eqref{eqcr} can be found in Appendix \ref{KdVsol} where it is shown that
 at least one root of the equation $d(\un)=0$ lies inside the interval $(0,2\tw^2)$ if $4\tw^2={v-\mu-\a b^2}={4w^2-\a b^2}>0$ and
\bal\la{kdvsing}
\text{Loop: }&
\left\{
\begin{array}{cc}
 b \neq 0\,, \quad 
\a < \a_-<0~ & \\
b \neq 0\,,\   4w^8>b^2\,,  \quad\ \ & \a_+^{(2)} <\a<\a_+^{(3)}<{4w^2\ov b^2}~
\end{array}
\right.
\\
\text{Bell or Double Loop: }&
\left\{
\begin{array}{cc}
 b = 0\,, \quad 
\alpha >\frac{27}{128 w^6} &~  \\
b \neq 0\,,\   4w^8>b^2\,, \ &\quad 0<\a_+^{(1)} <\a<\a_+^{(2)}
\end{array}
\right.
\eal
Here the critical values of $\a$ are given by
\bal\la{crval}
\a_-=-\frac{\sqrt{4 w^4+2|b|}-2 w^2}{b^2}\,,
\quad
\a_+^{(2)}=\frac{2 w^2-\sqrt{4 w^4-2 |b|}}{b^2}\,,
\quad \a_+^{(3)}=\frac{2 w^2+\sqrt{4 w^4-2 |b|}}{b^2}\,,
\eal
 and $\a_+^{(1)}$ is the positive root smaller than ${2 w^2\ov b^2}$ of the following equation
 \bal
 1-\frac{128}{27} \a \left( w^2-{\a b^2\ov8}\right)^3=0\,.
 \eal
 As one can see from \eqref{kdvsing}, the soliton solution is single-valued
 for 
\bal
\a_+^{(3)}<\a< \a_{\rm max}={4w^2\ov b^2}\,,\quad \frac{4w^8}{b^2}>1\,.
\eal
 It is interesting that this region is nonperturbative in $\a$. 

%\medskip

\begin{figure}[t!]
\centering
\includegraphics[width=0.32\linewidth]{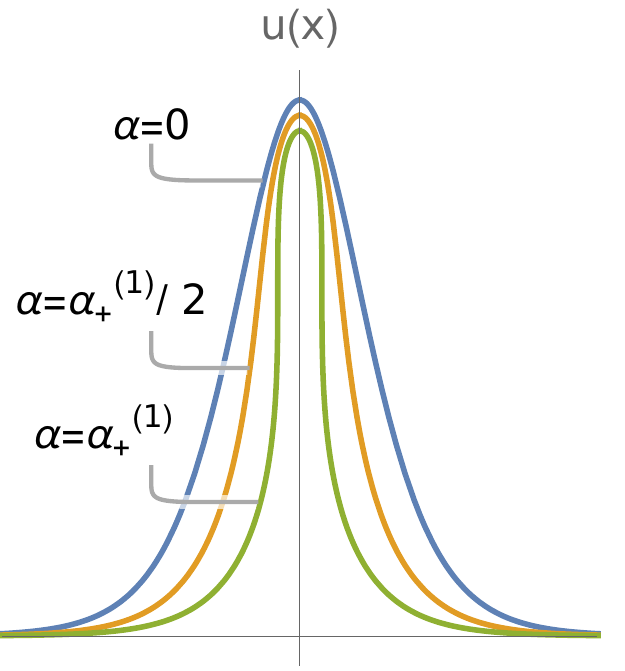}
\includegraphics[width=0.32\linewidth]{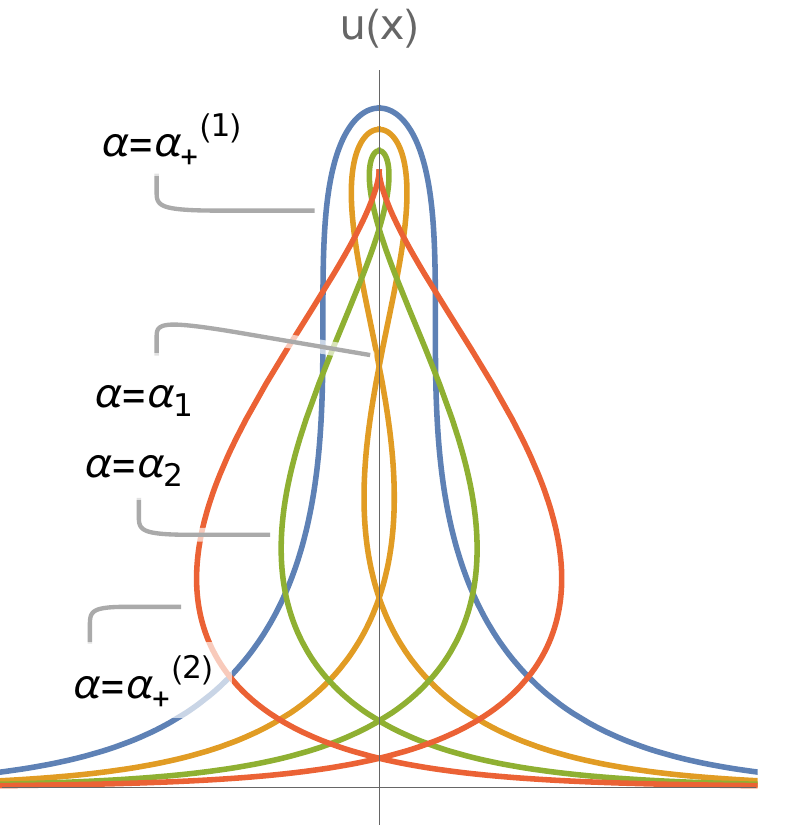}
\includegraphics[width=0.32\linewidth]{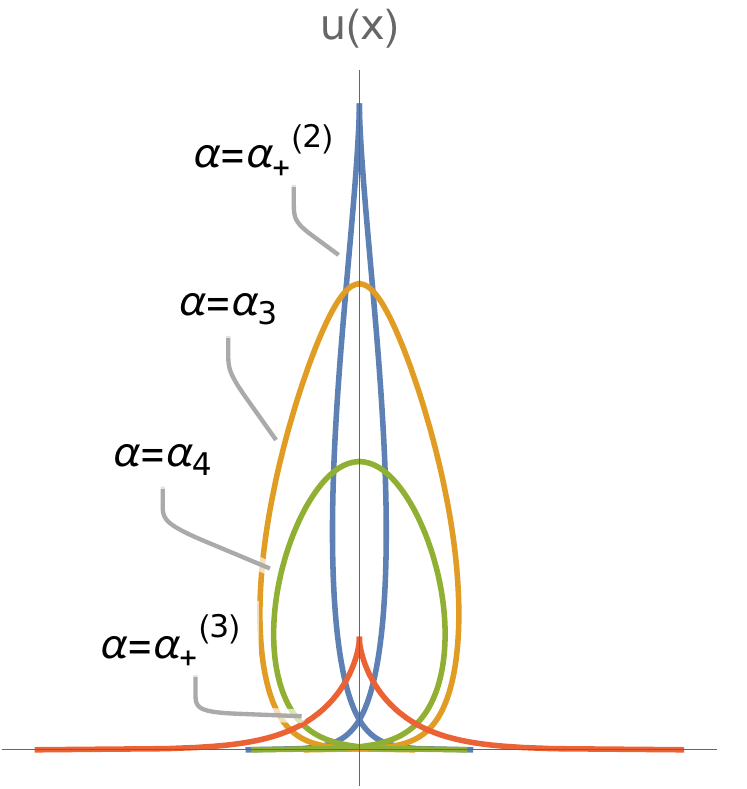}
\caption{Evolution of KdV soliton solutions for $w=1,\, b=1, \a > 0$, transitioning between different types of multi-valued solutions. Left: Width is decreasing with increasing $\a$, $\a_+^{(1)} \approx 0.23$. Centre: Formation of double-loop solution for $\a_+^{(1)} < \a < \a_+^{(2)}$, with a singular solution at $\a = \a_+^{(2)} \approx 0.59$. The intermediate values are equally spaced, $\a_1 = { 2\a_+^{(1)} + \a_+^{(2)}\ov3}$, $\a_2 = {\a_+^{(1)} + 2\a_+^{(2)}\ov3} $. Right: Amplitude decreasing, transitioning to singular peak at $\a = \a_+^{(3)} \approx 3.41$. $\a_3 = {2\a_+^{(2)} + \a_+^{(3)}\ov3}$, $\a_4 = {\a_+^{(2)} + 2\a_+^{(3)}\ov3}$.}
\label{fig:KdV1}
\end{figure}

\begin{figure}[h!]
\centering
\includegraphics[width=0.32\linewidth]{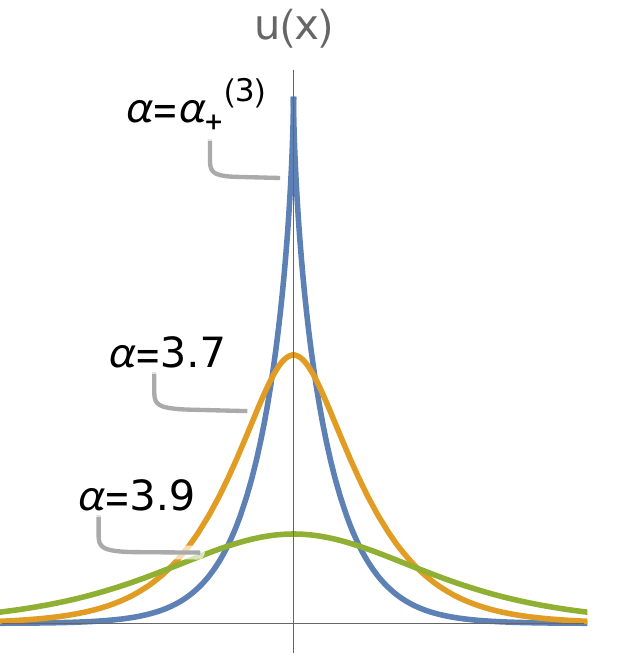}
\includegraphics[width=0.32\linewidth]{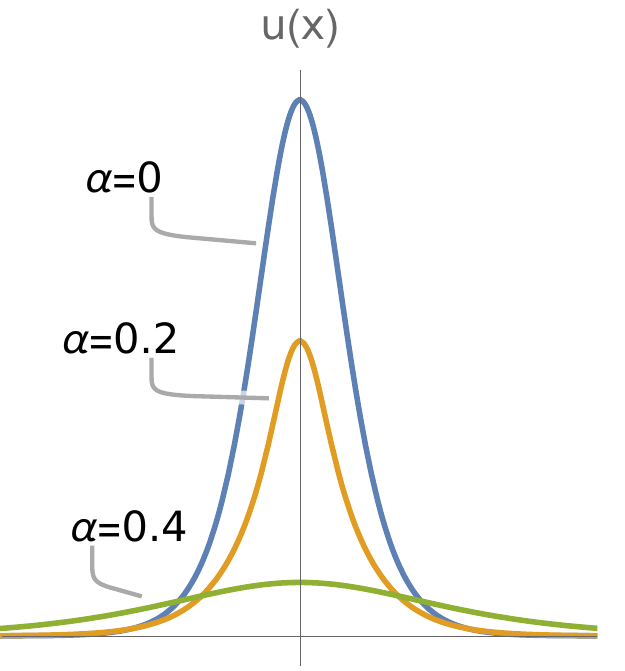}
\includegraphics[width=0.32\linewidth]{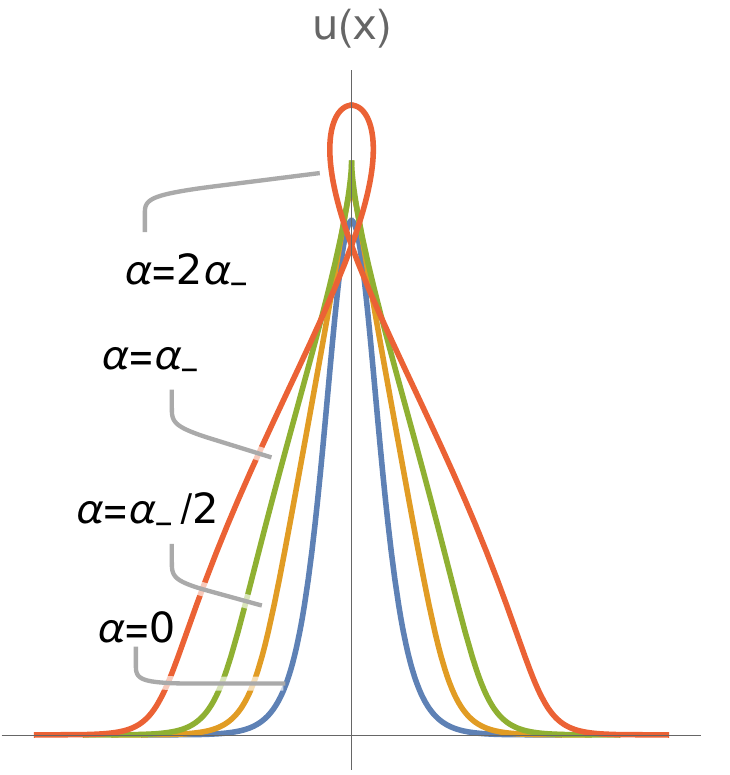}
\caption{Left: Continuation of evolution from figure \ref{fig:KdV1}, displaying single-valued solution for $\a > \a_+^{(3)} \approx 3.41$. The extreme flattening of the solution in the limit $\a \to 4$ is due to $\tw \to 0$. Centre: With $w=1, \, b=3$, the soliton remains regular for all $0 < \a < 4/9$, after which it ceases to exist in a similar fashion. Right: $w=1, \, b=1, \, \a < 0$, $\a_- \approx -4.4$. Solution widens, but with nonzero $b$ develops into a loop solution.}
\label{fig:KdV2}
\end{figure}

 The complex evolution of the solution for $\a > 0$ for which all $\a_+^{(i)}$ are real is shown in Figure \ref{fig:KdV1}. For large $\a$ the dominating factor is the $\a$ dependence in $\tw$ which enables the existence of the nonperturbative regular solutions for $\a > \a_+^{(3)}$.
These regular solutions are shown in Figure \ref{fig:KdV2}, along with the negative $\a$ behaviour.
The solution profiles for $b=0$ are shown in Figure \ref{fig:KdVb0} in Appendix \ref{KdVsol}. 
 
\medskip

Let us also mention that for $b=0$  the \TTb\ deformed soliton solution can be easily found by using the dynamical coordinate transformation \cite{Tateo18b}. We denote the coordinates of the undeformed soliton \eqref{solitonKdV} with $f(t)=0$ by $\tau\,,\, \s$, and its stress-energy tensor by $\cT^\g{}_\de$, and computing it on the soliton solution, we get
\bal
 \cT^\tau{}_\tau &=4 w^4 \text{sech}^4\left(w (\sigma -v\tau  )\right) \left(\mu -4w^2 +8
   w^2 \,\text{sech}^2\left(w (\sigma -v\tau  )\right)\right)\,,
   \quad
   \cT^\s{}_\tau &=v\,\cT^\tau{}_\tau \,,
   \\
   \cT^\tau{}_\s &=-4 w^4 \text{sech}^4\left(w (\sigma -v\tau  )\right)
   \,,\quad
   \cT^\s{}_\s =v\, \cT^\tau{}_\s\,.
\eal
The dynamical coordinate transformation is given by
\bal
dt&= (1+\a\,\cT^\s{}_\s )\,d\tau -\a\,\cT^\tau{}_\s \,d\s = d\tau -\a\,\cT^\tau{}_\s \, d(\s - v\tau)\,,
\\
 dx&= (1+\a\,\cT^\tau{}_\tau )\,d\s -\a\,\cT^\s{}_\tau \,d\tau=d\s +\a \,\cT^\tau{}_\tau\,d(\s-v\tau) \,,
 \\
 d(x-vt)&=\big(1 +\a \,(\cT^\tau{}_\tau+v\,\cT^\tau{}_\s)\big)\,d(\s-v\tau)\,.
\eal

 \begin{figure}[t!]
\centering
\includegraphics[width=0.35\linewidth]{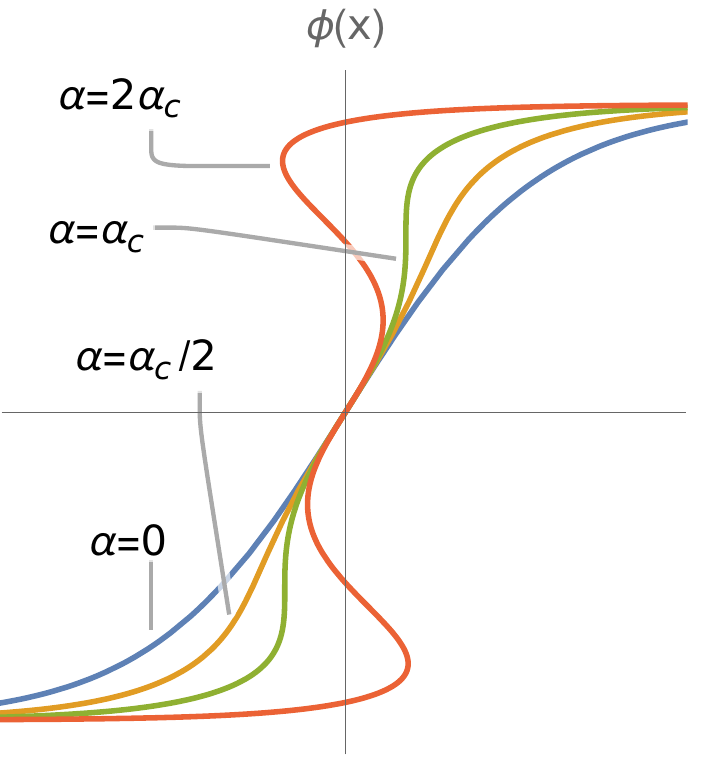}
\includegraphics[width=0.35\linewidth]{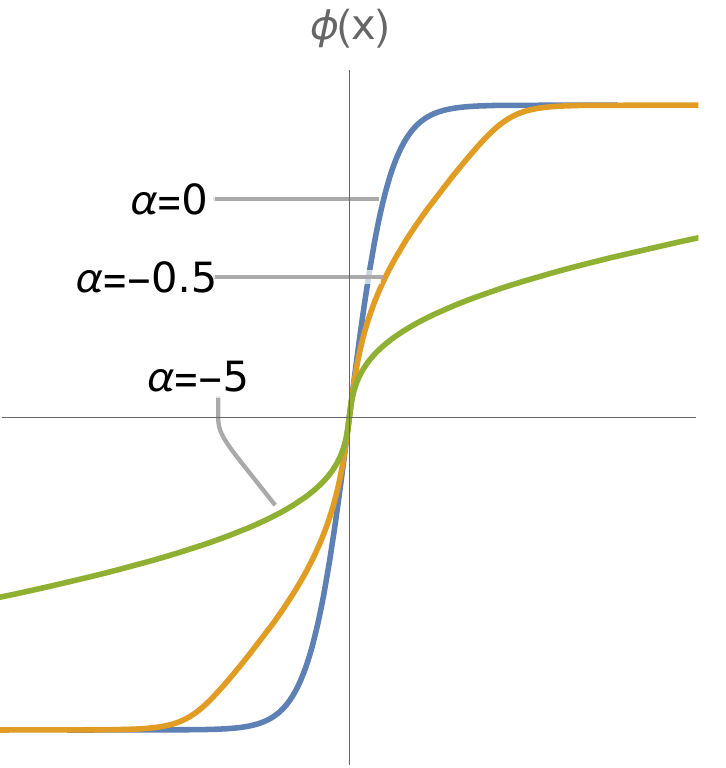}
\caption{Plots of the soliton $\phi(x-vt)$ with $w=1$. Note shock-wave behaviour is only displayed for $\a > \a_c = 27/128$.}
\label{fig:KdVsoliton}
\end{figure}

\noindent
Integrating this relation we find that the deformed inverse relation is
\bal
x-vt &= \s - v\tau + \frac{32 \a w^3}{15} \text{tanh}^3\left(w (\sigma -\tau  v) \right) \left(3\text{tanh}^2\left(w (\sigma -\tau  v) \right) -5 \right)\,,
\\
x-vt &= \frac{1}{w}\text{arctanh}\left(\frac{\p}{2 w}\right)+ \frac{\a \p^3}{15} \left(3 \p^2 - 20 w^2 \right)\,.
\eal
Note that $\p(x-vt) \in (-2w, 2w)$ and so the $\alpha$-dependent term has a fixed sign for all values of $w$. The deformed behaviour of the soliton is fixed by the sign of $\a$. By requiring the roots of $\frac{d(x-vt)}{d \phi}$ to be real and within the range of $\phi$ we find that the critical value of the deformed parameter is $\a_c = \frac{27}{128 w^6}$.  For $\a > \a_c$ the soliton becomes multi-valued as it transitions into a shock-wave solution. For all $\alpha < \a_c$ the soliton exists and becomes wider as $\a \to -\infty$. These behaviours are shown in figure \ref{fig:KdVsoliton}, and they are consistent with \cite{Cardy2020}. It is easy to check that this solution agrees with \eqref{xun} for $b=0$. One can also see that $\p$ exhibits a physical shock wave formation. Since $\p'$ develops singularities as $\a$ approaches $\a_{\rm c}$,  it cannot be identified with the  \TTb\ deformed KdV field.

%%%%%%%%%%%%%%%%%%%%%%%%%%%%%%%%%%%%%%%%%%
\subsection{Comments}\la{comm2}
%%%%%%%%%%%%%%%%%%%%%%%%%%%%%%%%%%%%%%%%%%

Here we summarise the main properties of one-soliton solutions of the \TTb\ deformed NLS and KdV models we have found in this section. 

\smallskip

A common property of the solitons is that 
their width appears to depend on the deformation parameter according to the general phenomenon of widening/narrowing the width of particles under the \TTb\ deformation \cite{Cardy2020}, see also \cite{Jiang2020a}.  However, contrary to the claim in \cite{Cardy2020},  whether  soliton's size is increasing or decreasing depends not only on the sign of the deformation parameter but also on the potential and soliton parameters. In the NLS case this more complicated behaviour is caused by the addition of  the time component (density of particles multiplied by the chemical potential $\mu$)  of the conserved U(1) current to the seed model. After the \TTb\ deformation this cannot be undone by a time dependent  U(1) transformation \eqref{u1q}, and leads to substantial changes in the soliton's properties. 
Clearly, the relativistic case is more restrictive because adding the time component of a conserved current breaks Lorentz invariance. It is also worth noting that in the absence of the chemical potential the width \eqref{fwhm} is increasing for negative $\a$ and decreasing for positive $\a$ which is opposite to what was observed in \cite{Cardy2020} and \cite{Jiang2020a}. This is explained by the fact that 
the energy \eqref{charges} of the NLS soliton is  given by $E=\frac{ P^2}{2m}-\frac{1}{24} g^4 m^3-\mu\,Q$, and for $\mu=0$ its  rest energy is negative. The existence  of the rest energy means that in the non-relativistic case the \TTb\ deformation is effectively a mixture of  the \TTb\ deformation with a stress-energy tensor shifted so that the rest energy is zero, and the $JP$ deformation discussed in \cite{Cardy2020} and \cite{Jiang2020a}. If the chemical potential is sufficiently negative then the width is 
widening or narrowing in accord with  \cite{Cardy2020}. 
In the KdV case with the parameter $b=0$ the width of the deformed soliton again behaves oppositely to \cite{Cardy2020} and \cite{Jiang2020a}. Since the rest energy of the  soliton  is zero, it is tempting to conclude that the effect of ``pure'' \TTb\ deformation is in fact opposite to what was observed in \cite{Cardy2020} and \cite{Jiang2020a} for the $JP$ deformation at least for models with solitons. 
In this respect it would be interesting to analyse
the  Bethe equations with the \TTb\ deformed S-matrix for the deformed NLS model in the attractive regime to see if the conclusions of \cite{Jiang2020a}  where the repulsive case was studied remain unchanged.

\smallskip

Another common property of the deformed solitons is that
for any values of the parameters of the solitons there is at least one critical value $\a_{\rm cr}$ at which solitons begin to exhibit the shock-wave behaviour. We proposed that for values of $\a$ beyond $\a_{\rm cr}$ a soliton solution may be constructed by gluing together the two branches of the soliton solution at the points where the first derivative of the soliton field diverges. Despite the divergency, the soliton energy and momentum are finite, and the dispersion relation is defined for all values of $\a$. A natural expectation is that the glued soliton is unstable, and it would be interesting to check it.

\smallskip

The \TTb\ deformed KdV  equation admits at least a one-parameter family of one-soliton solutions. The extra parameter $b$ can be introduced explicitly in the \TTb\ deformed Lagrangian by shifting the field $\p$ by $b t$, and requiring that $\p$ asymptotes to constants at space infinities. Then, $b$ 
 can be interpreted as the parameter of the deformation by the time component of the conserved current due to the invariance of the \TTb\ deformed Gardner model under constant shifts of $\p$. Since the parameter $b$ modifies the properties of the soliton, in particular, it appears in the dispersion relation, such an interpretation is probably the right one. It is however unclear to us why one has to impose constant space asymptotes on $\p$. If $b$ does not vanish then there is an upper bound on $\a$, and approaching the bound the soliton's amplitude decreases and finally vanishes.  Choosing properly other parameters of the soliton, one can make the bound negative. Thus, the parameter $b$ allows one to construct solutions which do not exist in the seed model.

%%%%%%%%%%%%%%%%%%%%%%%%%%%%%%%%%%%%%%%%%%
\section{Conclusions}
%%%%%%%%%%%%%%%%%%%%%%%%%%%%%%%%%%%%%%%%%%

In this paper we have explained in detail how the light-cone gauge approach to the \TTb\ deformation can be applied to non-Lorentz invariant models, and used it to derive the deformed Lagrangians of the three prominent non-relativistic models -- the nonlinear Schr\"odinger, the Landau-Lifshitz and the Gardner.
The \TTb\ deformed Lagrangians have been then used to find
 one-soliton solutions of the deformed NLS and KdV models. The properties of the Lagrangians and solitons have been discussed in the Comments subsections \ref{comm1},  \ref{comm2}, and here we discuss some 
of the many open questions to be addressed.

\smallskip

We have only considered the deformed models on a line.  It would be interesting to put the models on a circle and look for all possible solutions including those nonperturbative in $\a$ with  energy divergent in the limit $\a\to0$.  In fact, these solutions may exist even for Lorentz invariant models, e.g. for the \TTb\ deformed sigma model described by the Lagrangian \eqref{Lttbsigma}, see subsections \ref{comm1} for a detail discussion. 

\smallskip

The seed models we have considered are integrable, and it is believed that their \TTb\ deformations are integrable too. The first step in proving the integrability would be finding 
Lax pairs for the deformed models. Lax pairs of several models including the NLS model were recently found in \cite{Tian} by using the dynamical coordinate transformation \cite{Tateo18b}. Their results agree with the previously known Lax pairs of the sine-Gordon and Liouville models \cite{Tateo18a,Leoni}. It should be possible to apply the method of \cite{Tian}  to the matrix NLS model and the LL model. 
It would be interesting to see if their method can be generalised to include models of the Gardner type where auxiliary fields cannot be eliminated and one has to deal with them.

\smallskip

As has been mentioned in subsections \ref{comm1}, understanding the Poisson structure and developing a Hamiltonian formulation of the deformed models is important and  probably very hard. 

\smallskip

Given a Lax pair $(V,U)$ and a Hamiltonian formulation of the NLS model, one can calculate the Poisson bracket between $U$'s, and see how the $r$-matrix structure is modified, and whether it can be quantised. 

\smallskip

If a seed model possesses an additional conserved $U(1)$ current $J$ then  one can consider $JT$ deformations \cite{Guica17} which have properties similar to the \TTb\ deformation. 
The NLS model is one of the simplest nonrelativistic models with the $U(1)$ symmetry, and it would be interesting to analyse the properties of the model deformed by $JT$ operators. Some steps in this direction have been made in \cite{Jiang2020b,Tateo2020}. The light-cone gauge approach to the \TTb\ deformation of relativistic sigma models can be readily generalised to include the $JT$ deformations and deformations by  operators linear in conserved currents \cite{SF19b}. It should be possible to consider in the same framework nonrelativistic models. As was pointed out in \cite{Mezei2019a}, since the $JT$ deformations break Lorentz invariance  the deformations by operators linear in conserved currents are necessary to derive flow equations for the spectrum. 
 In fact, for nonrelativistic models it seems  necessary  to include the linear deformations even to derive the flow equations for the \TTb\ deformation. 
 
 \smallskip

The \TTb\ deformation of nonrelativistic models is defined with the help of the Lagrangian flow
$\pa_\a\cL =-\,$\TTb. This modifies the Poisson structure of a seed model, and makes it difficult to derive flow equations for the spectrum. It would be interesting to see whether one can define the deformation as the Hamiltonian flow  $\pa_\a\cH =\,$\TTb\  which preserves the Poisson structure of  a seed model. This can be done for a \TTb\ deformed massive Dirac fermion \cite{AAF} but for a bosonic model the Hamiltonian might appear to be nonlocal in space.

\smallskip 

Finally, there are many questions related to quantum \TTb\ deformed models. Some of them have been discussed in subsections \ref{comm1}, \ref{comm2}. 

%%%%%%%%%%%%%%%%%%%%%%%%%%%%%%%%%%
\appendix

%%%%%%%%%%%%%%%%%%%%%%%%%%%%%%%%%%
\section{Deformed NLS soliton solution}\la{NLSsol}

We start from the Lagrangian expressed in polar coordinates as in equation \ref{LNLS2} and derive the equations of motion. Then we apply the ansatz as described in equation \ref{eq:soliton_ansatz}. Furthermore we decompose $\r_A$ into real and imaginary components as $\r_A = X + i Y$.
In addition to the equations of motion for $(\r, \r_A, \bar\r_A)$ we have the following simplified equations from the continuity of the stress tensor and the fact that $\p$ is a cyclic variable
\bal
-v \frac{\partial \cL}{\partial \dot\p} + \frac{\partial \cL}{\partial \p'}=c_1\,, \quad  -v T^t_t + T^x_t = c_2\,, \quad -v T^t_x + T^x_x = c_3\,.\\
\eal
Applying the boundary conditions of $\r(\pm\infty) = 0$ to each of the equations of motion and continuity equations yields
\bal
X(\pm \infty) = Y{}(\pm \infty) = 0 = c_1 = c_2 = c_3\,.
\eal
Solving the equations of motion for $\p$ yields a simple relation for $Y$. Applying this to the continuity equations yields the relation for $X^2$ 
\bal
Y &= \frac{1}{2} v \r\,, \quad X^2 &= -\frac{1}{4} \r^2 \left(g^2 \r^2+4 \m +v^2-4 \omega\right)\,.
\eal
The two continuity equations for the stress tensor become dependent at this stage. From the real part of the equations of motion for $\r_A$ we find
\bal
X &= -\frac{1}{2} \r' \left(\a  g^2 \r^4+\a  v^2 \r^2+4 \a  \m  \r^2-2 \a  \omega \r^2-2\right)\,.
\eal
By substituting this into the continuity equation we find the first-order differential equation in $\r$ 
\bal
\r' = \pm\frac{2\r \sqrt{u^2 - g^2 \r^2}}{4-\a  \r^2 \left(2 g^2 \r^2+4 \m -u^2+v^2\right)}\,.
\eal
Where we redefine the arbitrary parameter $\omega = \frac{1}{4} \left(4 \m +u^2+v^2\right)$.
Then, if one considers the imaginary part of the equations of motion for $\r_A$ without substituting this new relation, we can find $\p'$ in terms of $\r, \r'$
\bal
\varphi' = \frac{1}{4} v \left(\a  \r'^2 \left(\a  \r^2 \left(2 g^2 \r^2+4 \m -u^2+v^2\right)-4\right)+2\right)\ \,.
\eal
Recalling that the integration variable is $x-vt$, one can trivially integrate the constant term to get $\varphi(\a = 0) = v/2(x-vt)$ and  use a change of coordinates $d(x-vt) = d\r (\r')^{-1}$. Then we find the expression for $\p$ in terms of $\r$
\bal
\phi &= \frac{t}{4} \left(4 \mu +u^2+v^2\right)+\frac{v}{2}(x-v t) \pm \frac{ \alpha  v \left(u^2-g^2 \r^2 \right)^{3/2}}{6 g^2}\,.
\eal
The auxiliary fields are
\bal
\r_A &= \frac{1}{2} \r \left(i v \pm \sqrt{u^2-g^2 \r^2}\right)\,,\quad \bar\r_A = \frac{1}{2} \r \left(-i v \pm \sqrt{u^2-g^2 \r^2}\right)\,.
\eal
The stress-energy tensor becomes
\bal
T^x_t & = -\frac{\r^2 \left(-2 g^2 \r^2-4 \mu +u^2+v^2\right)}{\alpha  \r^2 \left(2 g^2 \r^2+4 \mu -u^2+v^2\right)-4} = v T^t_t\,,\\
T^x_x & = \frac{2 v^2 \r^2}{\alpha  \r^2 \left(2 g^2 \r^2+4 \mu -u^2+v^2\right)-4} = v T^t_x\,.
\la{stress-eq:soliton_sol}
\eal

%\newpage
%%%%%%%%%%%%%%%%%%%%%%%%%%%%%%%%%%%%%%%%%%%%%%%%%%%%%%%%%%%%%%%%%%%%%%%%%%%%%%%%%%%
\section{Additional Graphs of NLS deformed soliton}\la{NLSfig}

\begin{figure}[h!]
    \centering
    \includegraphics[width = 0.32\linewidth]{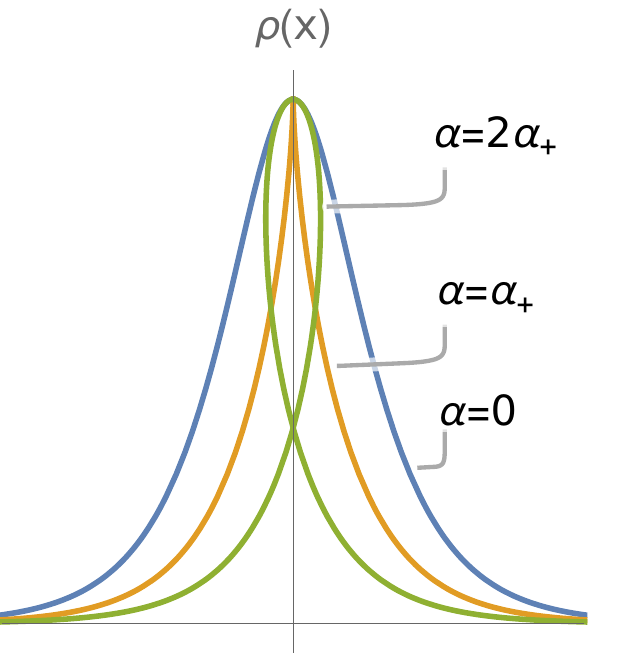}
    \includegraphics[width = 0.32\linewidth]{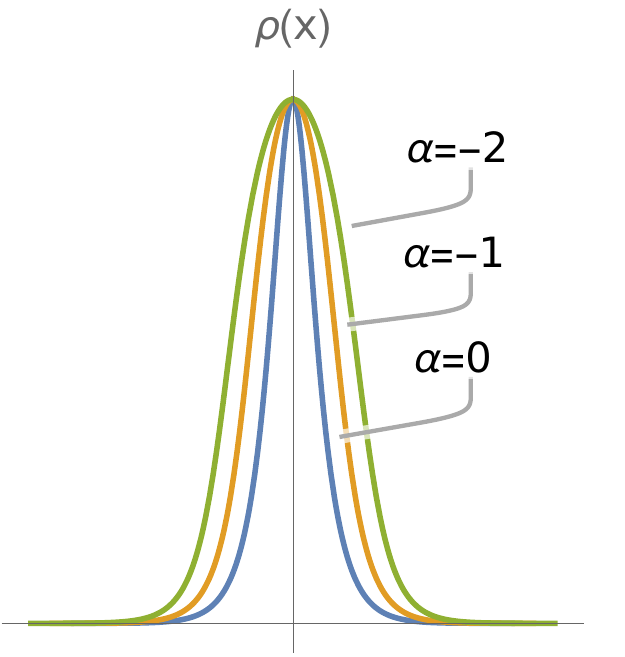}
     \includegraphics[width = 0.32\linewidth]{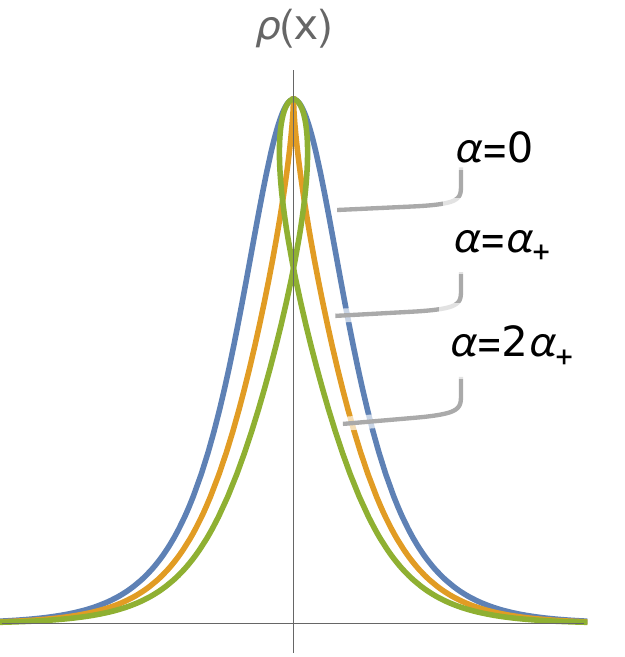}
    \caption{Left \& Centre: Case A, $\m = 1$, $\a_+ = 4/5$, displaying loop formation for $\a > \a_+ > 0$ and widening for $\a < 0$. Right: Case B, $\m = 0$, $\a_+ = 4$, loop solution appears for $\a > 0$, this is the only case with a finite region of valid $\a$.}
    \label{fig:NLSA}
\end{figure}

\begin{figure}[h!]
    \centering
     \includegraphics[width = 0.32\linewidth]{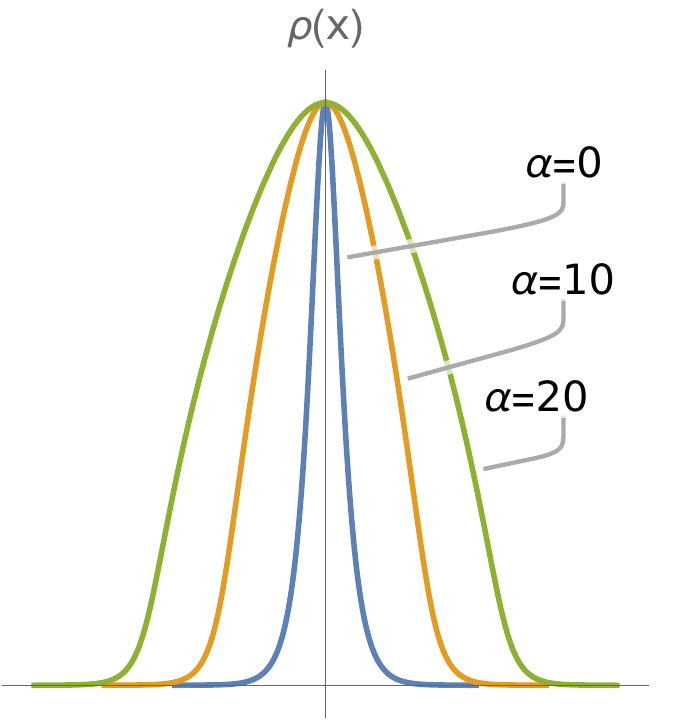}
    \includegraphics[width = 0.32\linewidth]{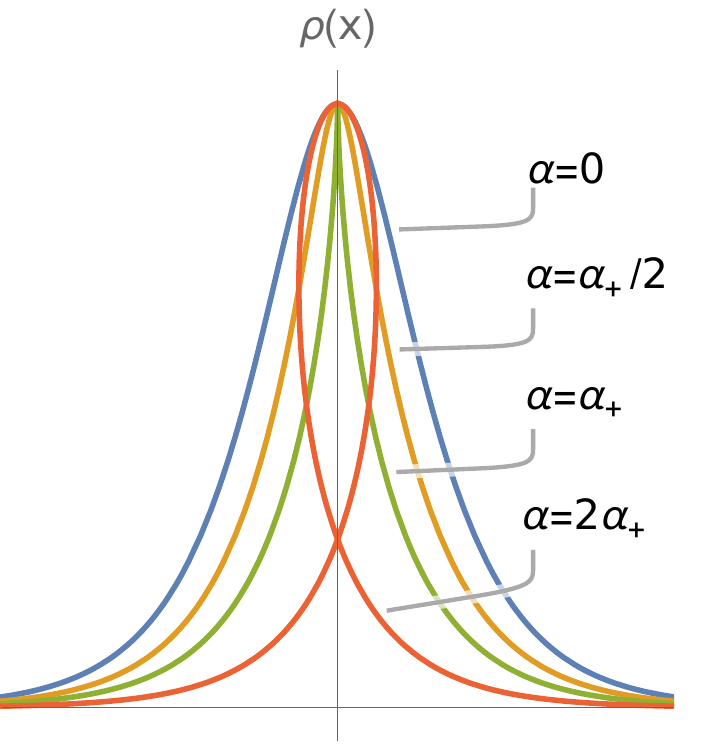}
    \includegraphics[width = 0.32\linewidth]{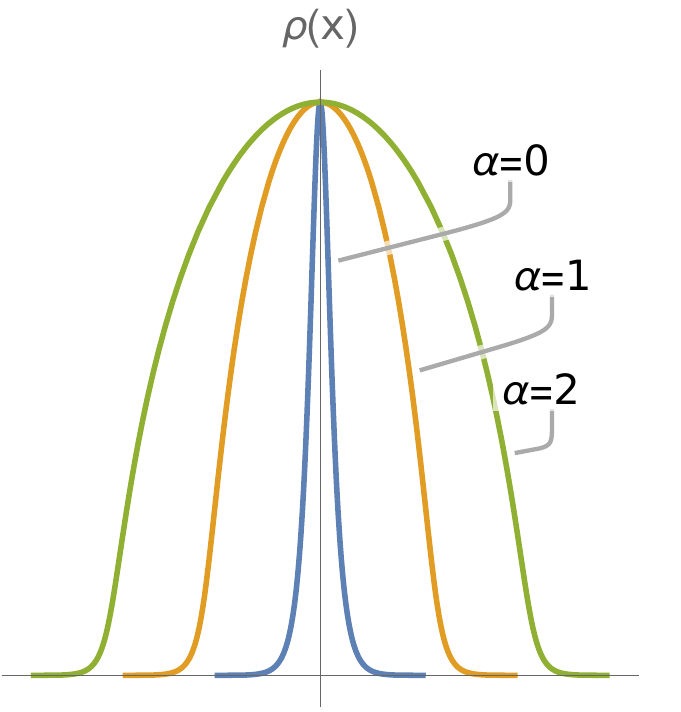}
     \caption{Left: Case C, $\m = -0.6$, $\a_- = -2.76817$, showing regular widening solution for $\a > 0$. Centre \& Right: Case D, $\m = -10$, $\a_+ = -4/39$. Loop formation for $\a < \a_+ < 0$, widening for $\a > 0$. Note the varying rate of soliton widening between the two cases.}
    \label{fig:NLSD}
\end{figure}

\newpage
%%%%%%%%%%%%%%%%%%%%%%%%%%%%%%%%%%%%%%%%%%%%%%%%%%%%%%%%%%%%%%%%%%%%%%%%%%%%%%%%%%%
\section{Deformed KdV soliton solution}\la{KdVsol}

The starting deformed Lagrangian is given by
\bal
\cL = \frac{-A \left(A+2 \un' \left(\a  \un \dot\p-1\right)-2 \a  \un \dot\un \p'\right)+B \left(\un-\p'\right)-\un \left(\un (2 \un+\m )+\dot\p\right)}{\a  A^2+\a  \un (\un (2 \un+\m )-B)+1}
\eal
In addition to the equations of motion for each of the fields, we use the simplified continuity equations for the stress tensor and the equation of motion for $\p$
\bal
-v \cJ^t + \cJ^x & = c_1 = \frac{B-v \un}{\a  A^2+\a  \un (\un (2 \un+\m )-B)+1}\,,\\
T^\s_\s - v T^\tau_\s & = c_2 = \frac{A^2+b \un-B \un+\m  \un^2+2 \un^3}{\a  A^2-\a  B \un+\a  \m  \un^2+2 \a  \un^3+1}\,,
\eal
where we have applied the ansatz given by \ref{KdV_ansatz}.

From the equations of motion for $\p$, we find an expression for $B$, which we substitute into the stress tensor continuity equation to solve for $A^2$
\bal
B &=\frac{\a  c_1 \left(A^2+\un^2 (2 \un+\m )\right)+v \un+c_1}{1+\a c_1 \un}\,, \\
A^2 &=\frac{\un (-\un (2 \un (\a c_2-1)-\a  b c_1+\a c_2 \m -\a c_2 v-\m +v)+b-c_1)-c_2}{\a c_2-1}\,.
\eal
Removing the $A^2$ in the solution for $B$, we then apply this to the equation of motion for $A$. We can then solve for $A$ and then create another equation by requiring the two solutions for $A$ be consistent
\bal
B &= \frac{-c_1+\un (\a  b c_1 + \a c_2 v-v)}{\a c_2-1}\,,\\
A &= \frac{\un' (\a  \un (2 \un (2 \un(\a c_2-1) -\a  b c_1+\a c_2 \m -\a c_2 v-\m +v)-b+2 c_1)+2 \a c_2-1)}{\a c_2+\a  c_1 \p'-1}\,.\\
\eal
At this stage we aim to fix the constants $c_1, c_2$ by evaluating the expressions as $x-vt \to \infty$. Initially we only have that $\p' \to 0$ in this limit, and the resulting expressions for the equations of motion for A, B, $\p$ and the consistency equation for the A solutions are nontrivial. However, the set of solutions for which these equations hold each require $\un =0$ and hence $\un'=0$ at infinity. With the new boundary conditions, we find that $c_2 = 0$. Applying the solutions and boundary conditions for the equation of motion for $\un$ then sets $c_1=b$. We find $\un'$ in terms of $\un, \p'$ by solving the consistency equation for the A solutions, and applying this to the equation of motion for B we then find the last relation for $\p'$ in terms of $\un$. 
\bal
\un' = \pm \frac{\un \left(\a  b \p'-1\right) \sqrt{ \left(v -\a  b^2-\m-2 \un \right)}}{\left(\a  \un \left(2 \un \left(2 \un+\a  b^2+\m -v\right)-b\right)+1\right)}\,, \quad \p' = \frac{\un-\a  b \un^2}{\a  \un^2 \left(4 \un+\a  b^2+2 \m -2 v\right)+1}\,.
\eal
Now we have expressed all the fields in terms of $\un$ and have a first-order differential equation for said field. The \TTb \  flow equation holds on shell, and the solutions hold in the undeformed limits $\a \to 0$ and $b \to 0$.
We can define the variable $\tw^2 = v - \m + b^2 \a =  w^2 + b^2 \a$ to simplify the expressions, where $w^2$ was used in the undeformed description of the soliton.
We can use a change of variables from $dx \to (\un')^{-1}d\un$ to perform spatial integration. The stress tensor on-shell is given by
\bal
T^\s_\s & = \frac{\un^2 \left(\a  b^2+\m +4 \tw^2\right)}{\a  \un^2 \left(-4 \un+\a  b^2+8 \tw^2\right)-1} = v T^\tau_\s\,,\\
T^\s_\tau & = \frac{\left(\m +4 \tw^2\right) \un^2 \left(-4 \un-\m +4 \tw^2\right)+b^2}{\a  \un^2 \left(-4 \un+\a  b^2+8 \tw^2\right)-1} = vT^\tau_\tau - b^2\,.\\
\eal
And the conserved current from the equations of motion is
\bal
\cJ^t = \frac{\un (\a  b \un+1)}{1-\a  \un^2 \left(-4 \un+\a  b^2+8 \tw^2\right)}\,.
\eal

\medskip

Let us now analyse the equations\eqref{eqcr} which we repeat here for convenience 
\bal\la{eqcra}
d(\un)\equiv \alpha  \un^2 \left(4 \un-8 \tw^2\right. -\a  b^2)+1 = 0, \qquad 0 < \un < 2 \tw^2\,,
\eal
 and determine the values of the parameters for which 
 the solution becomes multi-valued.
 
Calculating the values of $d(\un)$ at the boundaries of the allowed values of $\un$, one finds
\bal
d(0)=1\,,\quad d(2\tw^2) = 1- 4 \a^2b^2\tw^4= 1- {1\ov4}\a^2b^2(4w^2 - \a\,b^2)^2\le1\,.
\eal
Then, we find the first and second derivatives of $d(\un)$, and its extremal points
\bal
d'(\un)&=2 \a \un \left(6 \un+\a b^2-8 w^2\right)\,,\quad d''(\un)=2 \a \left(12 \un+\a b^2-8 w^2\right)\,,
\\
 \un^{\rm ex}_1=0\,,\quad d''(0)&=-2 \a \left(8 w^2-\a b^2\right)\,,\quad  \un^{\rm ex}_2={8 w^2-\a b^2\ov6}\,,\quad d''(\un^{\rm ex}_2)=2 \a \left(8 w^2-\a b^2\right)
\eal
Let us now fix $v,\mu,b$ and find for which values of $\a$ the equation \eqref{eqcra} has solutions in the interval $(0,2\tw)$.

\begin{figure}
\includegraphics[width=0.32\linewidth]{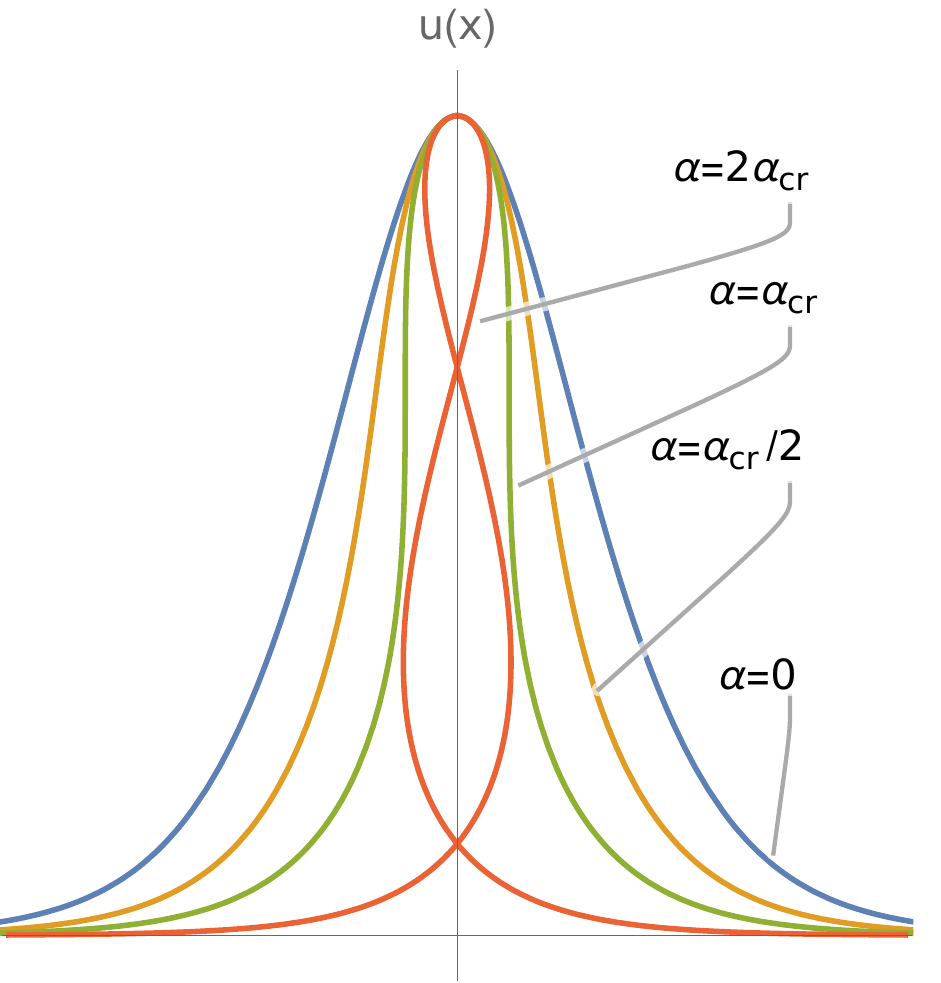}
\includegraphics[width=0.32\linewidth]{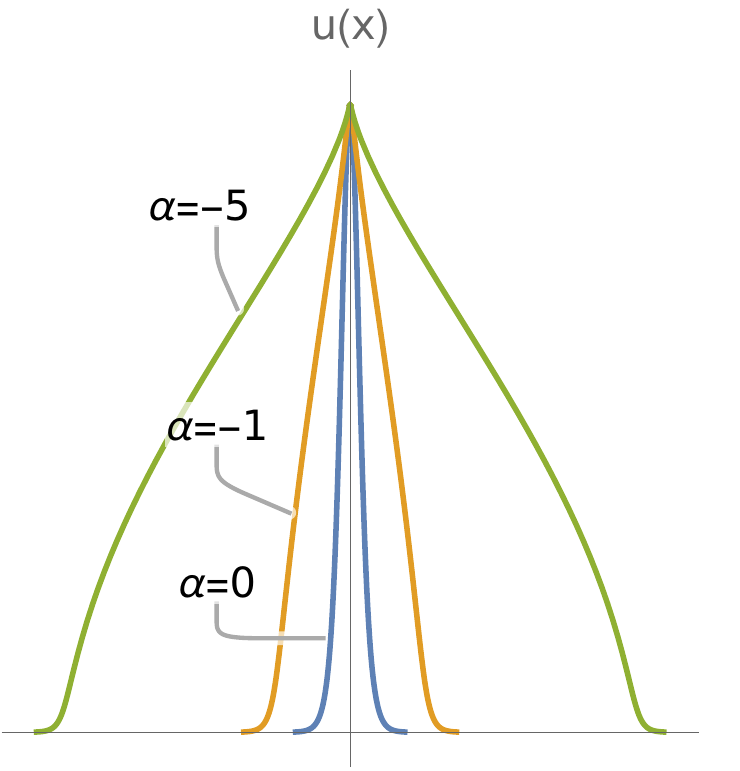}
\includegraphics[width=0.32\linewidth]{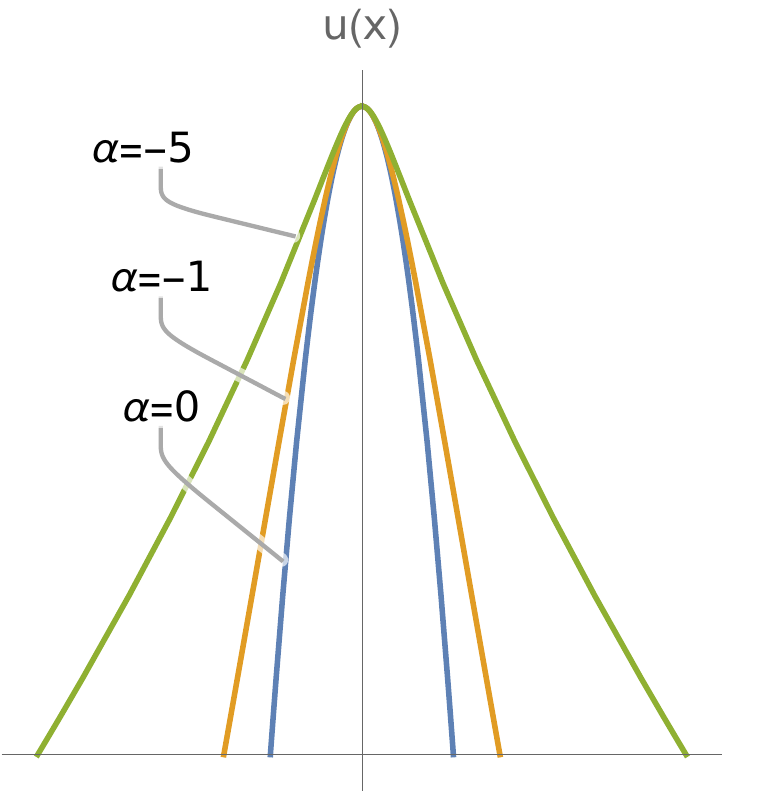}
\caption{KdV soliton solutions for $w=1,\, b=0$. Double-loop solution forms only for $\a > \a_c = 27/128$. For $\a < 0$, solution remains single-valued and increases in width. Rightmost plot examines peak of $\a < 0$ plot, indicating that the solution remains smooth at x=0. }
\label{fig:KdVb0}
\end{figure}

We begin with the simplest case $b=0$. Then, $\tw^2= w^2$ and
\bal
b=0:\quad d(0)= d(2\tw^2) = 1\,, \quad \un^{\rm ex}_2={4w^2\ov3}<2\tw^2\,,\quad d''(\un^{\rm ex}_2)=16 \a w^2\,.
\eal
Thus, the second extremal point is always inside the interval $(0,2\tw)$, and if $\a<0$ then it is a maximum and all roots of 
\eqref{eqcra} are outside the interval $(0,2\tw)$. If $\a>0$, then it is a minimum, and 
\bal
b=0:\quad d(\un^{\rm ex}_2)= 1-\frac{128 \a w^6}{27}\,.
\eal
It is clear now that for $b=0$ \eqref{eqcra} has two roots the interval $(0,2\tw)$ for $\a>\a_{\rm cr} = \frac{27}{128 w^6}$, and  the solution is first of a bell shape and then of a double loop shape as on the left picture of Figure \ref{fig:KdVb0}.

\medskip

If $b\neq0$ there may be critical values of $\a$ for both signs. 

\medskip

Let us first  consider the $\a<0$ case. 
If $w^2={v-\mu\ov4}>0$, then $8w^2-\a b^2>4w^2>0$ and
for all $\a<0$ the first extremal point $\un^{\rm ex}_1=0$ is a minimum, and  the second extremal point is a maximum. Therefore, one can get a root of  \eqref{eqcra}  which is inside the interval $(0,2\tw^2)$ only if $d(2\tw^2)$ becomes negative. 
Solving the equation $d(2\tw^2)=0$ with $\a<0$, one finds the negative critical value of $\a$
\bal\la{alpham}
\a_-=-\frac{\sqrt{4 w^4+2|b|}-2 w^2}{b^2}\,,
\eal
and for $\a<\a_-$  the solution is of a loop shape, see the right plot of Figure \ref{fig:KdV2}.

Then, if $w^2<0$ then $\a<{4w^2\ov b^2}<0$, and for ${8w^2\ov b^2}<\a<{4w^2\ov b^2}$ the first extremal point $\un^{\rm ex}_1=0$ is a maximum, and  the second extremal point $\un^{\rm ex}_2< 0$ is a minimum while for $\a<{8w^2\ov b^2}$,
$\un^{\rm ex}_1$ becomes a minimum, and  $\un^{\rm ex}_2> 0$ becomes a maximum. 
Thus, just as for the $w^2>0$ case one needs  $d(2\tw^2)$ to be negative which again happens at $\a=\a_-$ given by \eqref{alpham}. Depending on values of $w^2$ and $b$, $\a_-$ may be greater or less than $8w^2/b^2$.

\medskip

Let us now consider the $\a>0$ case. 
For all $\a>0$ the first extremal point $\un^{\rm ex}_1=0$ is a maximum, and therefore the second extremal point is a minimum, and
\bal
b\neq 0:\quad d(\un^{\rm ex}_2)=  1-\frac{128}{27} \a \left( w^2-{\a b^2\ov8}\right)^3\,.
\eal
As for the $b=0$ case, critical values are given by roots of the equation $d(\un^{\rm ex}_2)=0$. It is easy to see that $d(\un^{\rm ex}_2)$ as a function of $\a$ has the only minimum at 
\bal
 \a_{\rm min}={2 w^2\ov b^2}\quad\Rightarrow\quad d(\un^{\rm ex}_2)=  1-\frac{4w^8}{b^2} \,.
\eal
In fact, $\a_{\rm min}$ is also the minimum of $d(2\tw^2)$ and  $\un^{\rm ex}_2 = 2\tw^2=w^2$ for $\a=\a_{\rm min}$.

Thus, if $\frac{4w^8}{b^2}<1$ then there is no critical value for $\a>0$, and if $\frac{4w^8}{b^2}>1$ then there are two real roots of  the equation $d(\un^{\rm ex}_2)=0$, and the critical value $\a_+^{(1)}$ is the positive root which is smaller than $ \a_{\rm min}$. For values of $\a$ slightly greater than $\a_+^{(1)}$ there are two roots of the equation \eqref{eqcra} for values of $\un$ for which $\un'=\infty$, and therefore the solution
is first of  a bell shape and then of a double loop shape
as for $b=0$ case. The two roots are inside the interval $(0,2\tw^2)$ until $\a$ becomes equal to
\bal
\a_+^{(2)}=\frac{2 w^2-\sqrt{4 w^4-2 |b|}}{b^2}\,.
\eal
At $\a = \a_+^{(2)}$ one gets $d(2\tw^2)=0$ and therefore the larger root is equal to $2\tw^2$. Increasing $\a$ more moves the larger root away from the interval $(0,2\tw^2)$, and the solution is of a loop shape. Finally, the smaller root leaves the interval $(0,2\tw^2)$ at
\bal
\a_+^{(3)}=\frac{2 w^2+\sqrt{4 w^4-2 |b|}}{b^2}\,,
\eal
because $d(2\tw^2)$ is again equal to 0 for $\a = \a_+^{(3)}$. Thus, for
\bal
\a_+^{(3)}<\a< \a_{\rm max}={4w^2\ov b^2}\,,\quad \frac{4w^8}{b^2}>1\,,
\eal
the solution is regular again.

The discussion above is summarised in eq.\eqref{kdvsing}.

%%%%%%%%%%%%%%%%%%%%%%%%%%%%%%%%%%%%%%%

\end{document}